\documentclass[a4paper,11pt]{article}
\usepackage[dvips]{graphicx}
\usepackage{amsmath}
\usepackage{amssymb}
\usepackage{cite}
\usepackage{cancel}
\usepackage{fancybox}
\usepackage{ulem}
\usepackage{pifont}
\usepackage{colortbl}
\usepackage{multicol}

\usepackage{colortbl}
\usepackage{multicol}
\usepackage[all,knot]{xy}
\usepackage{longtable}

\makeatletter
 
 \@addtoreset{equation}{section}
\makeatother

\begin{document}

\begin{titlepage}

\begin{center}

\vspace{1cm}
{\Large\textbf{
$W$ and $Z$ Boson Pair Production at Electron-Positron Colliders\\ 
in Gauge-Higgs Unification
}}
\vspace{1cm}

\renewcommand{\thefootnote}{\fnsymbol{footnote}}
Naoki Yamatsu${}^{1}$\footnote[1]{yamatsu@phys.ntu.edu.tw},
Shuichiro Funatsu${}^{2}$,
Hisaki Hatanaka${}^{3}$,\\
Yutaka Hosotani${}^{4}$,
and
Yuta Orikasa${}^{5}$
\vspace{5mm}

\textit{
$^1${Department of Physics, National Taiwan University, Taipei, Taiwan 10617, R.O.C.}\\
$^2${Ushiku, Ibaraki 300-1234, Japan } \\
$^3${Osaka, Osaka 536-0014, Japan} \\
$^4${Research Center for Nuclear Physics, Osaka University, Ibaraki,
 Osaka 567-0047, Japan}
$^5${Institute of Experimental and Applied Physics, Czech Technical
 University in Prague, \\ 
Husova 240/5, 110 00 Prague 1, Czech Republic}
}

\date{\today}

\abstract{
 We examine $W$ and $Z$ boson pair production  processes 
 at electron-positron collider experiments
 in the $SU(3)_C\times SO(5)_W\times  U(1)_X$  gauge-Higgs
 unification (GHU) model.
 We find that the deviation of the total cross section for the
 $e^-e^+\to W^-W^+$ process from the Standard  Model (SM) in the GHU
 model with parameter sets, which are consistent with the current
 experiments,
 is about 0.5\% to 1.5\% and 0.6\% to 2.2\% for $\sqrt{s}=250$\,GeV and
 500\,GeV, respectively, depending on the initial electron and  positron
 polarization.
 We find that for the $e^-e^+\to ZZ$ process the deviation from the
 SM in the GHU model is at most 1\%. 
 We find that unitality bound for the $e^-e^+\to W^-W^+$ process is
 satisfied in the GHU model as in the SM, as a consequence of the
 relationship among coupling constants.
}

\end{center}
\end{titlepage}



\section{Introduction}
\label{Sec:Introduction}

The Standard Model (SM) in particle physics has been established at low
energies. However, it is not yet clear whether the observed Higgs boson
has exactly the same properties as those in the SM. 
The Higgs couplings to quarks, leptons, and SM gauge bosons, as well as
the Higgs self-coupling, need to be determined more accurately in future
collider experiments such as the International Linear Collider
(ILC)\cite{Behnke:2013xla,Baer:2013cma,Adolphsen:2013jya,Adolphsen:2013kya,Behnke:2013lya,ILCInternationalDevelopmentTeam:2022izu}, 
the Compact Linear Collider (CLIC)\cite{Linssen:2012hp}, the Future
Circular Collider (FCC-ee)\cite{Blondel:2021ema},
the Cool Copper Collider (C${}^3$) \cite{Dasu:2022nux},  Circular
Electron Positron Collider  (CEPC)\cite{CEPCStudyGroup:2018ghi}, and
muon collider \cite{Accettura:2023ked}.

The SM Higgs boson sector has many problems, one of which is the fact
that there are large corrections to the Higgs boson mass at the quantum
level. 
Fine-tuning of the bare mass is required to obtain the observed Higgs mass
$m_h=125.25\pm 0.17$\,GeV \cite{ParticleDataGroup:2022pth}. One known
way to 
stabilize the mass of the Higgs boson against quantum corrections is to
identify the Higgs boson as a zero mode of the 5th dimensional
component of the gauge potential. This scenario is called gauge-Higgs
unification (GHU)
\cite{Hosotani:1983xw,Hosotani:1988bm,Davies:1987ei,Davies:1988wt,Hatanaka:1998yp,Hatanaka:1999sx}.
In GHU models, the Higgs boson appears as a fluctuating mode in the
Aharonov-Bohm (AB) phase $\theta_H$ in the 5th dimension; 
$SU(3)_C \times SO(5)_W\times U(1)_X$ GHU models in the Randall-Sundrum
(RS) warped space have been proposed in 
Refs.~\cite{Agashe:2004rs,Medina:2007hz,Hosotani:2008tx,Funatsu:2019xwr},
where $SO(5)_W\supset SU(2)_L\times SU(2)_R$.
The GHU models are classified into two types, depending on
whether quarks and leptons belong to the vector or the spinor
representation of $SO(5)_W$. 
The GHU model whose quarks and leptons belong to the spinor
representation of $SO(5)_W$ can be regarded as a low-energy effective
description of the $SO(11)$ GHU model
\cite{Hosotani:2015hoa,Yamatsu:2015rge,Furui:2016owe,Hosotani:2017edv,Hosotani:2017ghg,Englert:2019xhz,Englert:2020eep},
where the SM gauge symmetry $SU(3)_C\times SU(2)_L\times U(1)_Y$ is
embedded in the $SO(11)$ grand unified gauge symmetry
\cite{Georgi:1974sy,Inoue:1977qd,Fritzsch:1974nn,Gursey:1975ki,Slansky:1981yr,Yamatsu:2015gut}
in higher dimensional framework
\cite{Kawamura:1999nj,Kawamura:2000ir,Kawamura:2000ev,Hall:2001pg,Hall:2001zb,Burdman:2002se,Lim:2007jv,Kojima:2011ad,Kojima:2016fvv,Kojima:2017qbt,Yamatsu:2017sgu,Yamatsu:2017ssg,Yamatsu:2018fsg,Maru:2019lit,Kawamura:2022ecd}.
The phenomena of the GHU model below the electroweak (EW) scale are very
close to those of the SM in a parameter regime that satisfies the
current experimental constraints on the Kaluza-Klein (KK) mass 
$m_{\rm KK}\gtrsim13$\,TeV and the AB phase $\theta_H \lesssim0.1$
\cite{Funatsu:2019xwr,Funatsu:2019fry,Funatsu:2020znj,Funatsu:2020haj,Funatsu:2021gnh,Funatsu:2021yrh,Funatsu:2022spb,Funatsu:2023jng,Hosotani:2023yhs}.
The strongest constraints come from the Large Hadron Collider (LHC)
experiment at $\sqrt{s}=13$\,TeV with up to 140\,fb${}^{-1}$ data
\cite{Aad:2019fac,ATLAS:2020yat,ATLAS:2019lsy,ATLAS:2019fgd,CMS:2019gwf,CMS:2021ctt}
by using the $Z'$ and $W'$ boson search results for the $pp\to\ell\nu$
and $pp\to\ell^-\ell^+$ processes \cite{Funatsu:2021yrh},
where the $Z'$ bosons are mixed vector bosons of $U(1)_X$,
$U(1)_L(\subset SU(2)_L)$, and $U(1)_R(\subset SU(2)_R)$ and 
the $W'$ bosons are mixed vector
bosons of $SO(5)_W/(U(1)_L\times U(1)_R)$.

In the future $e^-e^+$ collider experiments, it is possible to explore
up to the region of tens of TeV in terms of the KK mass scale
\cite{Funatsu:2017nfm,Yoon:2018xud,Yoon:2018vsc,Funatsu:2019ujy,Funatsu:2020haj,Funatsu:2022spb,Funatsu:2023jng,Bilokin:2017lco,Richard:2018zhl,Irles:2019xny,Irles:2020gjh,Fujii:2017vwa,Aihara:2019gcq,Bambade:2019fyw,Poschl:2022ecb,Irles:2023ojs,Irles:2023nee}.
Large parity violation appears in the coupling of quarks and leptons to
KK gauge bosons, especially to the first KK modes; 
the sign of the bulk mass parameter of the fermions in the GHU model is
an important factor in determining whether 
the coupling constants of the $Z'$ and $W'$ bosons to the right- or
left-handed fermions are larger.
We studied observables such as cross sections and asymmetries
\cite{Schrempp:1987zy,Kennedy:1988rt,Abe:1994wx,Abe:1996nj} in the
process of fermion pair production. 
Due to the very large cross section of the processes, we can clearly
observe deviations from the SM in the early stage of the ILC experiment
$(\sqrt{s}=250$\,GeV, integral luminosity $L_{\rm int}=250$\,fb$^{-1}$)
even when $m_{\rm KK}$ is larger than the current experimental
constraint $m_{\rm KK}\simeq13$\,TeV.
The cross sections of the processes are very sensitive to the initial
polarizations of the electron and the positron, so the sign of the
corresponding bulk mass of each fermion in each final state in the GHU
model can also be determined by analyzing the polarization dependence.
Recently, we examined the single Higgs production processes such as
the Higgs strahlung process $e^-e^+\to Zh$ \cite{Funatsu:2023jng}. 
By using the Higgs strahlung process, it is possible to explore up to
the region of tens of TeV in terms of the KK mass scale.

The vector boson pair production processes at $e^-e^+$ collider are
significant.
Measurements of the $W^\pm$ boson pair production process near the
threshold are important for
determining the $W$ boson mass $m_W$. 
The deviation from the SM prediction has been recently reported
by the Collider Detector at Fermilab
(CDF) Collaboration at the Tevatron \cite{CDF:2022hxs}.
In the SM, the contributions to the $e^-e^+\to W^-W^+$ process come from
the s-channel process via $\gamma$ and $Z$ boson and the t-channel via
neutrinos.
In s-channel processes and t-channel processes alone, a factor
$s/m_W^2$ appears in the cross section from the longitudinal
polarization vector of the $W$ boson in the final state.
The unitality bound on the high energy behavior of
the total cross section is given by
$\sigma_{\rm total}\leq C\{\mbox{log}(s)\}^2$, 
known as the Froissart bound
\cite{Froissart:1961ux,Langacker:2017uah}, where $C$ is a constant.
For $s\gg m_W^2$, the total cross section may behave in such a way that
unitality is violated because each contribution from the s-channel
and t-channel are $O(s)$.
In the SM, individual cross sections that would break unitality are
miraculously canceled out by the cross sections between them because
special conditions are satisfied between the coupling constants.
As a result, the total cross section of $e^-e^+\to W^-W^+$, which
includes both s- and t-channel contributions, satisfies Froissart
(unitality) bound. This is required by the Goldstone boson equivalence
theorem \cite{Cornwall:1974km,Chanowitz:1985hj,Peskin:1995ev}, where 
this theorem was first proofed in Ref.~\cite{Cornwall:1974km}. The proof
of the theorem is based on the Ward identities of the spontaneously
broken gauge theory.
In Ref.~\cite{Funatsu:2016uvi}, $W^\pm$ boson pair production has been
partially analyzed in the LHC experiment, but not in the $e^-e^+$
collider, and the GHU model discussed in Ref.~\cite{Funatsu:2016uvi} is 
different from the GHU model discussed in the paper.

In this paper we analyze the $W$ and $Z$ boson pair production processes
$e^-e^+\to W^-W^+$ and $e^-e^+\to ZZ$ in the GHU model to clarify the
difference between the predictions in the SM and the GHU model.
We show that unitality bound for the $e^-e^+\to W^-W^+$ process
is satisfied in the GHU model as well as in the
SM by investigating the asymptotic behavior of the cross sections for
large $\sqrt{s}$.
We calculate the energy and angle dependence of the cross sections and
clarify the differences between the predictions of the SM and GHU models.
We show that the $e^-e^+\to W^-W^+$ process with 
$(\sqrt{s}\,,L_{\rm int})=$(250\,GeV,\,1\,ab$^{-1}$),
(500\,GeV,\,2\,ab$^{-1}$) at the ILC can explore up to the region of
tens of TeV in terms of the KK mass scale, which is beyond the current
constraints on the KK mass scale from the LHC experiment. 
We analyze the $e^-e^+\to ZZ$ process in the same way and show that the
deviation from the SM is at most 1\%.

The paper is organized as follows.
In Sec.~\ref{Sec:Model}, the $SU(3)_C\times SO(5)_W\times U(1)_X$ GHU
model is introduced. 
In Sec.~\ref{Sec:Parameter-sets},
we give some parameter sets of the GHU model.
In Sec.~\ref{Sec:Cross-section}, 
we give the formulas for the cross sections of the 
$e^-e^+\to W^-W^+$ and $e^-e^+\to ZZ$ processes,
involving the $Z'$ and $W'$ bosons as well as the $Z$ and $W$ bosons.
We show that unitality bound in the $e^-e^+\to W^-W^+$ process
is satisfied in the GHU model.
In Sec.~\ref{Sec:Results},
we present numerical results for the cross sections
of $e^-e^+\to W^-W^+$ and $e^-e^+\to ZZ$.
Section~\ref{Sec:Summary} is devoted to summary and discussions.

\section{Model}
\label{Sec:Model}

In this paper, we focus on observables related with the  EW gauge bosons
and leptons at tree level. The $SU(3)_C$ gauge bosons and fermions 
except leptons are not directly involved, so we omit them. For the full
field content in the GHU model, see Ref.~\cite{Funatsu:2019xwr}, 
in which the $SU(3)_C \times SO(5)_W\times U(1)_X$ GHU model was
originally proposed.

The GHU model is defined in the RS warped space with the following
\cite{Randall:1999ee}:
\begin{align}
 ds^2= g_{MN} dx^M dx^N =e^{-2\sigma(y)} \eta_{\mu\nu}dx^\mu dx^\nu+dy^2,
\end{align} 
where $M,N=0,1,2,3,5$, $\mu,\nu=0,1,2,3$, $y=x^5$,
$\eta_{\mu\nu}=\mbox{diag}(-1,+1,+1,+1)$,
$\sigma(y)=\sigma(y+ 2L)=\sigma(-y)$,
and $\sigma(y)=ky$ for $0 \le y \le L$.
By using the conformal coordinate $z=e^{ky}$
($1\leq z\leq z_L=e^{kL}$) in the region $0 \leq y \leq L$,
the metric is rewritten by 
\begin{align}
ds^2= \frac{1}{z^2}
\bigg(\eta_{\mu\nu}dx^{\mu} dx^{\nu} + \frac{dz^2}{k^2}\bigg).
\end{align} 
The bulk region $0<y<L$ ($1<z<z_L$) is anti-de Sitter (AdS) spacetime 
with a cosmological constant $\Lambda=-6k^2$, which is sandwiched by the
UV brane at $y=0$ ($z=1$) and the IR brane at $y=L$ ($z=z_L$).  
The KK mass scale is $m_{\rm KK}=\pi k/(z_L-1)$.

The $SO(5)_W\times U(1)_X$ symmetry includes the EW symmetry
$SU(2)_L\times U(1)_Y$, where $SO(5)_W\supset SU(2)_L \times SU(2)_R$.
$A_M^{SO(5)_W}$ and $A_M^{U(1)_X}$ represent
the $SO(5)_W$ and $U(1)_X$ gauge fields, respectively. 
The orbifold boundary conditions (BCs) $P_j(j=0,1)$ 
of the gauge fields on the UV brane $(y=0)$ and the IR brane $(y=L)$ are
given by 
\begin{align}
&\begin{pmatrix} A_\mu \cr  A_{y} \end{pmatrix} (x,y_j-y) =
P_{j} \begin{pmatrix} A_\mu \cr  - A_{y} \end{pmatrix} (x,y_j+y)P_{j}^{-1}
\label{Eq:BC-gauge}
\end{align}
for each gauge field, where $(y_0, y_1) = (0, L)$. 
For the $U(1)_X$ gauge boson $A_M^{U(1)_X}$, $P_0=P_1=1$.
For the $SO(5)_W$ gauge boson $A_M^{SO(5)_W}$,
$P_0=P_1=P_{\bf 5}^{SO(5)_W}$,
where $P_{\bf 5}^{SO(5)_W}=\mbox{diag}\left(I_{4},-I_{1}\right)$.
The orbifold BCs of the $SO(5)_W$ symmetry break
$SO(5)_W$ to $SO(4)_W\simeq SU(2)_L \times SU(2)_R$.
$W$, $Z$ bosons and $\gamma$ (photon) are zero modes in the 
$SO(5)_W\times U(1)_X$ of 4 dimensional (4D) gauge bosons, whereas the
4D Higgs boson is a zero mode in the $SO(5)_W/SO(4)_W$ part of the 5th
dimensional gauge boson.
In the GHU model, extra neutral gauge bosons $Z'$ correspond to
the KK photons $\gamma^{(n)}$, the  KK $Z$ bosons $Z^{(n)}$,
and the KK $Z_R$ bosons $Z_R^{(n)}$ ($n \ge 1$),
where the $\gamma$, and $Z$, $Z_R$ bosons are the mass eigen states of
the electro-magnetic (EM) $U(1)_{\rm EM}$ neutral gauge bosons of
$SU(2)_L$, $SU(2)_R$, and $U(1)_X$.
Extra charged gauge bosons $W'$ correspond to
the KK $W$ boson $W^{\pm(n)}$ ($n \ge 1$) and  the KK $W_R$ bosons
$W_R^{\pm(n)}$ ($n \ge 1$).

The SM lepton multiples are identified with the zero modes
of the lepton multiplets $\Psi_{({\bf 1,4})}^{\alpha}$ $(\alpha=1,2,3)$
in the bulk, where the subscript of $\Psi_{({\bf 1,4})}^{\alpha}$ stands
for the representations of $SU(3)_C\times SO(5)_W$.
The bulk fields $\Psi_{({\bf 1,4})}^{\alpha}$ obey the following BCs:
\begin{align}
\Psi_{({\bf 1,4})}^{\alpha} (x, y_j - y) = 
 - P_{\bf 4}^{SO(5)_W} \gamma^5 \Psi_{({\bf 1,4})}^{\alpha} (x, y_j + y),
\label{leptonBC1}
\end{align}
where $P_{\bf 4}^{SO(5)_W}=\mbox{diag}\left(I_{2},-I_{2}\right)$.
From the BCs in Eq.~(\ref{leptonBC1}), the left-handed Weyl fermion
in $({\bf 2,1})$ of $SU(2)_L\times SU(2)_R$ and 
the right-handed Weyl fermion in $({\bf 1,2})$ of 
$SU(2)_L\times SU(2)_R$ have zero mode.
Note that fermions in $({\bf 1,1})(0)$ of 
$SU(3)_C\times SO(5)_W\times U(1)_X$ are also introduced as the brane
fermions 
on the UV brane to reproduce
tiny neutrino masses via the seesaw mechanism in the GHU model
\cite{Hosotani:2017ghg}.

The brane scalar field in ${\bf 4}$ of $SO(5)_W$ is introduced to
realize the EW symmetry $SU(2)_L\times U(1)_Y$ at the EW scale.
A spinor {\bf 4} of $SO(5)_W$ is decomposed into
$({\bf 2}, {\bf 1}) \oplus ({\bf 1}, {\bf 2})$ of
$SO(4)_W \simeq SU(2)_L \times SU(2)_R$.
We assume that the brane scalar develops a nonvanishing vacuum
expectation value (VEV), which reduces the symmetry 
$SO(4)_W \times U(1)_X$ to the EW gauge symmetry 
$SU(2)_L\times U(1)_Y$, and the VEV of the brane scalar is much larger
than $m_{\rm KK}$ to ensure that the orbifold BCs for the 4D components
of the $SU(2)_R \times U(1)_X/U(1)_Y$ gauge fields become effectively
Dirichlet conditions at the UV brane \cite{Furui:2016owe}.

The $U(1)_Y$ gauge boson is realized as a linear combination of
$U(1)_R(\subset SU(2)_R)$ and $U(1)_X$ gauge bosons. The $U(1)_Y$ gauge
field $B_M^Y$ is given in terms of the $SU(2)_R$ gauge fields $A_M^{a_R}$
$(a_R=1_R,2_R,3_R)$ and the $U(1)_X$ gauge field $B_M$ by 
$B_M^Y = \sin\phi A_M^{3_R} + \cos\phi  B_M$.
Here the mixing angle $\phi$ between $U(1)_R$ and $U(1)_X$ is given by 
$\cos \phi= {g_A}/{\sqrt{g_A^2+g_B^2}}$ and
$\sin \phi= {g_B}/{\sqrt{g_A^2+g_B^2}}$, where
$g_A$ and $g_B$ are gauge coupling constants in $SO(5)_W$ and $U(1)_X$,
respectively.
The 4D $SU(2)_L$ gauge coupling constant is given by $g_w=g_A/\sqrt{L}$.
The 5 dimensional (5D) gauge coupling constant $g_Y^{\rm 5D}$ of
$U(1)_{Y}$ and the 4D 
bare Weinberg angle at the tree level, $\theta_W^0$, are given by
\begin{align}
&g_Y^{\rm 5D} =\frac{g_Ag_B}{\sqrt{g_A^2+g_B^2}}, \ \ \
\sin \theta_W^0 = \frac{\sin\phi}{\sqrt{{1 +\sin^2\phi}}}.
\label{Eq:gY-sW}
\end{align}

The 4D Higgs boson $\phi_H(x)$ is the zero mode contained in the 
$A_z = (kz)^{-1} A_y$ component:
\begin{align}
A_z^{(j5)}(x,z)= \frac{1}{\sqrt{k}} \, \phi_j (x) u_H (z) + \cdots,\
u_H (z) = \sqrt{ \frac{2}{z_L^2 -1} } \, z,\ \ 
\phi_H(x) = \frac{1}{\sqrt{2}} \begin{pmatrix} \phi_2 + i \phi_1 \cr \phi_4 - i\phi_3 \end{pmatrix} .
\end{align}
We take $\langle \phi_1 \rangle, \langle \phi_2 \rangle, 
\langle \phi_3 \rangle  =0$ and  
$\langle \phi_4 \rangle \not= 0$, which is 
related to the AB phase $\theta_H$ in the fifth dimension by
$\langle \phi_4 \rangle  = \theta_H f_H$, where
$f_H  = 2g_w^{-1}k^{1/2}L^{-1/2}(z_L^2 -1)^{-1/2}$.

We will give a part of the bulk action below, where the full action is
given in Ref.~\cite{Funatsu:2019xwr}.
The action of each gauge field, $A_M^{SO(5)_W}$ or $A_M^{U(1)_X}$,
is given in the form 
\begin{align}
S_{\rm bulk}^{\rm EW\, gauge}&=
\int d^5x\sqrt{-\det G}\, \bigg[-\mbox{tr}\left(
\frac{1}{4}F_{}^{MN}F_{MN}
+\frac{1}{2\xi}(f_{\rm gf})^2+{\cal L}_{\rm gh}\right)\bigg],
\label{Eq:Action-bulk-gauge}
\end{align}
where $\sqrt{-\det G}=1/k z^5$, $z=e^{ky}$,
$\mbox{tr}$ is a trace over all group generators for each group,
and 
$F_{MN}$ is a field strength defined by  
$F_{MN}:=\partial_MA_N-\partial_NA_M-i g[A_M,A_N]$
with each 5D gauge coupling constant $g$.
The second and third terms in Eq.~(\ref{Eq:Action-bulk-gauge})
are the gauge fixing term and the ghost term given in
Ref.~\cite{Funatsu:2019xwr},
respectively.
The action for the lepton sector in the bulk is given by
\begin{align}
&S_{\rm bulk}^{\rm lepton} =  \int d^5x\sqrt{-\det G} \,
\sum_{\alpha=1}^3
\overline{\Psi_{({\bf 1,4})}^{\alpha}}  
\left[
\gamma^A {e_A}^M
\bigg( D_M+\frac{1}{8}\omega_{MBC}[\gamma^B,\gamma^C]  \bigg) 
-c_L^\alpha\sigma'(y)
\right]
\Psi_{({\bf 1,4})}^{\alpha},
\label{fermionAction}
\end{align} 
where $\overline{\Psi_{({\bf 1,4})}^{\alpha}}=
i \Psi_{({\bf 1,4})}^{\alpha}{}^\dag\gamma^0$,
$\sigma'(y):=d\sigma(y)/dy$ and $\sigma'(y) =k$ for $0< y < L$. 
Each lepton multiplet $\Psi_{({\bf 1,4})}^{\alpha}(x,y)$ has
a bulk mass parameter $c_L^\alpha$ $(\alpha=1,2,3)$.

\section{Parameter sets}
\label{Sec:Parameter-sets}

To evaluate cross sections and other observable quantities in $W$ and
$Z$ boson pair production processes $e^-e^+\to W^-W^+$ and 
$e^-e^+\to ZZ$  at tree level in the GHU model, we need to know the
masses, decay widths, and coupling constants of the gauge bosons,
and the leptons. 
Parameters of the model are determined in the steps described in
Refs.~\cite{Funatsu:2020haj,Funatsu:2023jng}.

We present several parameter sets of the coupling constants of the
leptons, which are necessary for the present analysis, where
we omit quantities such as the mass and the decay width of $Z^{(1)}$
boson in the GHU model shown in Refs.~\cite{Funatsu:2023jng}.
In Sec.~\ref{Sec:Model}, we gave only the 5D Lagrangian of the GHU model,
but in the analysis, we use a 4D effective theory
with KK mode expansion of the 5th dimension.
By solving the equations of motions derived from the BCs of each 5D
multiplet, we can obtain the mass spectra of 4D modes for each 5D
field. Once mass spectra of a field are known, wave functions of the
zero mode and the KK modes of the field can be determined by substituting 
mass spectra into the mode function of the field.
Furthermore, coupling constants can be obtained by performing
overlap integrals of the wave functions of the corresponding fields.
For more details, see Ref.~\cite{Funatsu:2023jng}.

We will describe the steps to fix parameter sets in the GHU model,
where we will show parameter sets for leptons, the gauge bosons, and the
Higgs boson.
\begin{enumerate}
 \item We pick the values of $\theta_H$ and
       $m_{\rm KK}=  \pi k (z_L-1)^{-1}$.
       From the constraints on $\theta_H$ and $m_{\rm KK}$ from 
       the LHC-Run 2 results in the GHU model \cite{Funatsu:2021yrh}, we
       only consider parameters satisfying
       $\theta_H\leq 0.10$ and $m_{\rm KK}\geq 13$\,TeV.

 \item $k$ is determined in order for the $Z$ boson mass $m_Z$ to be
       reproduced, which fixes the warped factor $z_L$ as well.
       (For the mass formula of $Z$ boson, see
       Ref.~\cite{Funatsu:2019xwr}.) 

 \item The bare Weinberg angle $\theta_W^0$ in the GHU model is given 
       in Eq.~(\ref{Eq:gY-sW}).
       For each value of  $\theta_H$, the value of $\theta_W^0$ 
       is determined self-consistently to fit the observed 
       forward-backward asymmetry at $Z$ pole as in
       Ref.~\cite{Funatsu:2023jng}.

\begin{table}[htb]
{
\begin{center}
\begin{tabular}{c|cc|cc|cccc|c}
\hline
 \rowcolor[gray]{0.9}
 &&&&&&&&&\\[-0.75em]
\rowcolor[gray]{0.9}
 Name&$\theta_H$&$m_{\rm KK}$&$z_L$&$k$
 &$m_{\gamma^{(1)}}$&$\Gamma_{\gamma^{(1)}}$
 &$m_{\nu^{(1)}}$&$m_{e^{(1)}}$
 &$\sin^2\theta_W^0$\\
\rowcolor[gray]{0.9}
 &\mbox{[rad.]}&[TeV]&&[GeV]&[TeV]&[TeV]&[TeV]&[TeV]&\\ 
\hline 
 &&&&&&&&\\[-0.75em]
 A$_{-}$&0.10&13.00&
 3.865$\times10^{11}$&1.599$\times10^{15}$&10.198&3.252&13.039&13.039&0.2306\\
 A$_{+}$&0.10&13.00&
 4.029$\times10^{11}$&1.667$\times10^{15}$&10.198&3.256&13.034&13.034&0.2318\\
 B$_{-}$&0.07&19.00&
 1.420$\times10^{12}$&8.589$\times10^{15}$&14.887&4.951&18.852&18.852&0.2309\\
 B$_{+}$&0.07&19.00&
 1.452$\times10^{12}$&8.779$\times10^{15}$&14.887&4.951&18.848&18.848&0.2315\\
 C$_{-}$&0.05&25.00&
 5.546$\times10^{10}$&4.413$\times10^{14}$&19.649&5.862&25.528&25.528&0.2310\\
 C$_{+}$&0.05&25.00&
 5.600$\times10^{10}$&4.456$\times10^{14}$&19.649&5.864&25.527&25.527&0.2313\\
\hline
\end{tabular}
 \caption{\small
 The name of the parameter set and the corresponding $z_L$, $k$, and
 $\sin^2\theta_W^0$ for each $\theta_H$ and $m_{\rm KK}$
 are listed.
 In the SM, $\sin^2\theta_W(\overline{\mbox{MS}})=0.23122\pm0.00004$ 
 at $Z$ pole
\cite{ParticleDataGroup:2022pth}.
 The column ``Name'' denotes each parameter set.
 The subscripts of $A_\pm$, $B_\pm$, $C_\pm$ denote the sign of the
 bulk masses of the leptons. For example, $A_{+}$ denotes the case where
 the bulk mass of the lepton is positive and $A_{-}$ denotes the case
 where the bulk mass of the lepton is negative.
 The names of the parameter sets are the same as those in
 Ref.~\cite{Funatsu:2023jng}.
 }
\label{Table:Parameter-sets}
\end{center}
}
\end{table}

       The parameter sets of $(\theta_H,m_{\rm KK})$, named
       $A_{\pm}$,$B_{\pm}$ and $C_{\pm}$, used in this
       analysis are summarized in Table~\ref{Table:Parameter-sets},
       where the subscripts denote the sign of the  bulk masses of the
       leptons. For example, $A_{+}$ denotes the case where the bulk mass
       of the lepton is positive and $A_{-}$ denotes the case
       where the bulk mass of the lepton is negative.

 \item With given $\sin \theta_W^0$, wave functions of the gauge bosons
       are fixed. 
       Masses and widths of $\gamma^{(1)}$ is listed for each
       parameter set in Table~\ref{Table:Parameter-sets}, and
       those of $Z^{(1)}$, $Z_R^{(1)}$, $W$, $W^{(1)}$, $W_R^{(1)}$
       bosons are listed for each  parameter set in
       Table~4 in Ref.~\cite{Funatsu:2023jng}.

 \item The bulk masses of the leptons $c_L^\alpha$ in
       Eq.~(\ref{fermionAction})
       are determined so as to reproduce the masses of charged leptons,
       as the same in Ref.~\cite{Funatsu:2023jng}.
       The bulk masses and the brane interaction
       parameters of the leptons are listed in
       Table~5 in Ref.~\cite{Funatsu:2023jng}.

 \item With given the bulk masses and the brane interaction parameters,
       wave functions of fermions are fixed. 

 \item The mass of the Higgs boson can be obtained from the effective
       potential of the Higgs boson
       \cite{Funatsu:2020znj}.
       The mass of the Higgs boson is determined by adjusting the
       bulk mass of the dark fermions so that the mass of the Higgs
       boson is  $m_h=125.25\pm 0.17$\,GeV
       \cite{ParticleDataGroup:2022pth}.

\end{enumerate}

\begin{table}[htb]
\begin{center}
\begin{tabular}{c|cccc}
\hline
 \rowcolor[gray]{0.9}
 &&&&\\[-0.75em]
\rowcolor[gray]{0.9}
 Name
 &$g_{\gamma^{(1)}ee}^L$&$g_{\gamma^{(1)}ee}^R$
 &$g_{Zee^{(1)}}^L$&$g_{Zee^{(1)}}^R$
 \\
 \hline
 A$_{-}$
 &$-2.75873$&$+0.10788$&$-0.01346$&$0$
 \\
 A$_{+}$
 &$+0.10727$&$-2.54746$&$-0.01347$&$0$
 \\
 B$_{-}$
 &$-2.81496$&$+0.10455$&$-0.00951$&$0$
 \\
 B$_{+}$
 &$+0.10464$&$-2.81951$&$-0.00951$&$0$
 \\
 C$_{-}$
 &$-2.57731$&$+0.11144$&$-0.00664$&$0$
 \\
 C$_{+}$
 &$+0.11149$&$-2.67947$&$-0.00665$&$0$
 \\
\hline
\end{tabular}
 \caption{\small
 Coupling constants of $\gamma^{(1)}$ boson to electrons and
 $Z$ boson to electron and 1st KK electron
 in units of $g_w=e/\sin\theta_W^0$
 are listed.
 When the value is less than $10^{-5}$, we write $0$.
 }
\label{Table:Gauge-Charged-Lepton-Couplings}
\end{center}
\end{table}

\begin{table}[htb]
\begin{center}
\begin{tabular}{c|cc|cccc}
\hline
 \rowcolor[gray]{0.9}
 &&&&&&\\[-0.75em]
\rowcolor[gray]{0.9}
 Name
 &$m_{\nu_{\rm MeV}}$&$m_{\nu_{\rm MeV}'}$
 &$g_{We\nu_{\rm MeV}e}^L$&$g_{We\nu_{\rm MeV}e}^R$
 &$g_{We\nu_{\rm MeV}'e}^L$&$g_{We\nu_{\rm MeV}'e}^R$\\
 \rowcolor[gray]{0.9}
 &[MeV]&[MeV]&&&&\\
 \hline
 A$_{+}$
 &$9.735$&$9.735$
 &$-0.03534$&$-0.00177$&$+0.03534$&$+0.00177$\\
 B$_{+}$
 &$13.904$&$13.904$
 &$-0.02474$&$-0.00087$&$+0.02474$&$+0.00087$\\
 C$_{+}$
 &$19.464$&$19.464$
 &$-0.01768$&$-0.00044$&$+0.01768$&$+0.00044$\\
\hline
\end{tabular}
 \caption{\small
 MeV scale neutrino masses and coupling constants of $W$ boson to
 electron are listed  in units of  MeV and $g_w/\sqrt{2}$, respectively.
 There are no such neutrinos for negative lepton bulk masses.
 $\Delta m_{\nu_{\rm MeV}}:=
 m_{\nu_{\rm MeV}}-m_{\nu_{\rm MeV}'}\simeq
 4.0\times 10^{-7}$\,MeV,
 $8.2\times 10^{-7}$\,MeV, $1.6\times 10^{-6}$\,MeV for
 $A_+$, $B_+$, $C_+$, respectively.
 }
\label{Table:MeV-Neutrino-Gauge-Couplings}
\end{center}
\end{table}

\begin{table}[htb]
\begin{center}
\begin{tabular}{c|cccc}
\hline
 \rowcolor[gray]{0.9}
 &&&&\\[-0.75em]
\rowcolor[gray]{0.9}
 Name
 &$g_{\gamma^{(1)}WW}$
 &$g_{ZWW}$
 &$g_{Z^{(1)}WW}$
 &$g_{Z_R^{(1)}WW}$
 \\
\rowcolor[gray]{0.9}
 &[GeV]
 &[GeV]
 &[GeV]
 &[GeV]
 \\
 \hline
 A$_{-}$
 &$-0.00029$&$+0.87715$&$-0.00019$&$+0.00027$
 \\
 A$_{+}$
 &$-0.00029$&$+0.87647$&$-0.00019$&$+0.00027$
 \\
 B$_{-}$
 &$-0.00014$&$+0.87698$&$-0.00009$&$+0.00013$
 \\
 B$_{+}$
 &$-0.00014$&$+0.87664$&$-0.00009$&$+0.00013$
 \\
 C$_{-}$
 &$-0.00008$&$+0.87693$&$-0.00005$&$+0.00007$
 \\
 C$_{+}$
 &$-0.00008$&$+0.87676$&$-0.00005$&$+0.00007$
 \\
\hline
\end{tabular}
 \caption{\small
 Coupling constants of neutral gauge bosons to $W$ boson in units of
 $g_w$ are listed.
 In the SM and the GHU, $g_{\gamma WW}=e$, and 
 $g_{\gamma WW}/g_w=0.48085$.
 In the SM, $g_{ZWW}/g_w=\cos\theta_W=0.87680$.
 }
\label{Table:Triple-Gauge-Couplings}
\end{center}
\end{table}

Next, coupling constants of the gauge bosons and the leptons are obtained
from the five-dimensional gauge interaction terms by substituting the
wave functions of gauge bosons and fermions and integrating over the
fifth-dimensional
coordinate\cite{Funatsu:2014fda,Funatsu:2015xba,Funatsu:2016uvi}. 
The coupling constants of the gauge bosons to the leptons are obtained
by performing overlap integrals of the wave functions in the fifth
dimension of gauge bosons and leptons. Coupling constants of
$\gamma^{(1)}$ to electron are listed in
Table~\ref{Table:Gauge-Charged-Lepton-Couplings}. Coupling constants of
gauge bosons to leptons are listed in Table~7 in
Ref.~\cite{Funatsu:2023jng} for $W$ boson to leptons Tables~7 and 8 in
Ref.~\cite{Funatsu:2023jng} for $Z$ boson to neutrinos and charged
leptons.
MeV scale neutrino masses and coupling constants of $W$ boson to
electron are listed  in Table~\ref{Table:MeV-Neutrino-Gauge-Couplings},
where MeV neutrinos appear only when the bulk mass of the lepton is
positive.
The coupling constants of neutral gauge bosons to $W$ boson are obtained
by performing overlap integrals of the wave functions in the fifth
dimension of triple gauge bosons. Coupling constants of neutral gauge
bosons to $W$ boson pair are listed in
Table~\ref{Table:Triple-Gauge-Couplings}.

\section{Cross section}
\label{Sec:Cross-section}

In this section, we give the formulas necessary to calculate
observables of the $e^-e^+\to W^-W^+$ and $e^-e^+\to ZZ$ processes.
In the SM, the cross sections of the $e^-e^+\to W^-W^+$ and 
$e^-e^+\to ZZ$ processes are calculated in e.g.,
Refs.~\cite{Alles:1976qv,Brown:1978mq,Brown:1979ux,Frixione:1992pj}.
We can use Mathematica and its package FeynCalc
\cite{Mertig:1990an,Shtabovenko:2016sxi,Shtabovenko:2020gxv} in this
section for computational checks, such as the contractions of the square of the amplitude.

\subsection{Cross section of $e^-e^+\to W^-W^+$}

\subsubsection{Amplitude}

\begin{figure}[htb]
\begin{center}
\includegraphics[bb=0 0 512 247,height=3cm]{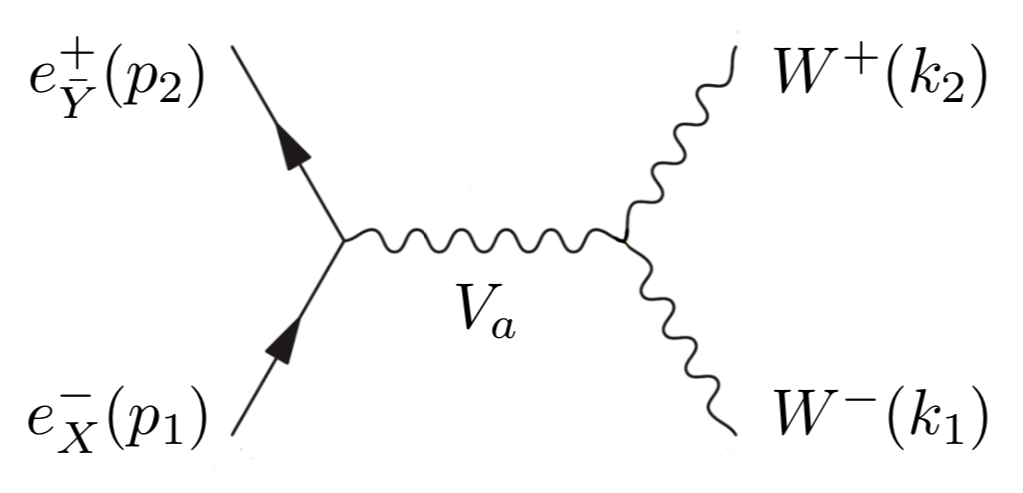}
\includegraphics[bb=0 0 430 228,height=3cm]{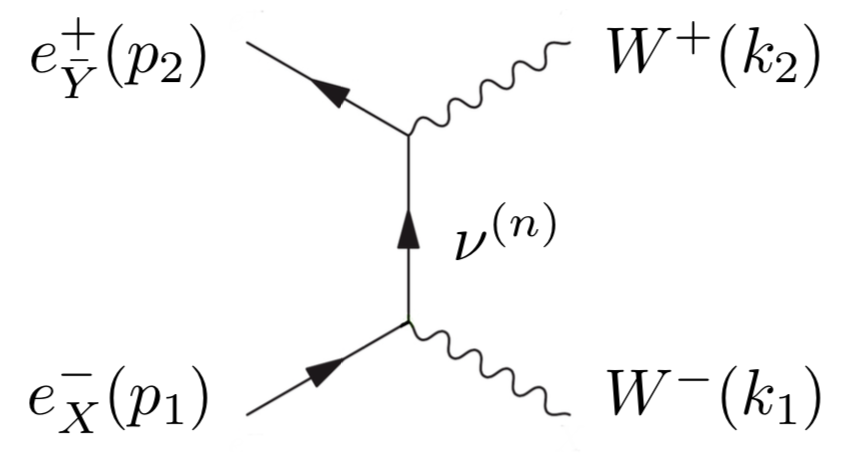}
 \caption{\small
 The Feynman diagrams for the  s-channel and t-channel
 contributions to the $e^-e^+\to W^-W^+$ process are shown.
 For the s-channel contribution, 
 $V_a$ stands for $\gamma,Z$ in the SM and for
 $\gamma^{(n)},Z^{(n)},Z_R^{(n)}$ $(n=1,2,...)$ in the GHU model.
 For the t-channel contribution, 
 $\nu^{(n)}$ stands for $\nu_e$ in the SM and for
 $\nu_{e}^{(n)}$ (and $\nu_{e2}^{(n)}$) in the GHU model.
 }
 \label{Figure:ee-to-WW}
\end{center}
\end{figure}

We consider the $W^-W^+$ production processes:
\begin{align}
e_X^-(p_1) e_{\bar{Y}}^+(p_2) \stackrel{}{\longrightarrow}
 W^-(k_1) W^+(k_2),
\end{align}
where $X,Y=L,R$; $p_1$ and $p_2$ are the momenta of the initial states
of electron and positron; $k_1$ and $k_2$ are the momenta of the final
states of $W^-$ and $W^+$ bosons. 
This is shown in Figure~\ref{Figure:ee-to-WW}.
For massive bosons $W^\pm$ and
massless electrons $e^{\pm}$, in the center-of-mass frame, we use the
following basis: 
\begin{align}
p_1^\mu &= \frac{\sqrt{s}}{2}(1,0,0,+1),
\ \ \
p_2^\mu = \frac{\sqrt{s}}{2}(1,0,0,-1),
\nonumber\\
k_1^\mu &= \frac{\sqrt{s}}{2}
\left(1,+\beta_W\sin\theta,0,+\beta_W\cos\theta\right),
\ \ \
k_2^\mu = \frac{\sqrt{s}}{2}
\left(1,-\beta_W\sin\theta,0,-\beta_W\cos\theta\right),
\label{Eq:kinematics}
\end{align}
where $\beta_W:=\sqrt{1-\frac{4m_W^2}{s}}$.
The Mandelstam variables are defined as
\begin{align}
&s:=(p_1+p_2) ^2=(k_1+k_2)^2,
\nonumber\\
&t:=(p_1-k_1)^2=(p_2-k_2)^2=m_W^2-\frac{s}{2}(1-\beta_W\cos\theta),
\nonumber\\
&u:=(p_1-k_2)^2=(p_2-k_1)^2
=m_W^2-\frac{s}{2}(1+\beta_W\cos\theta),
\end{align}
where $s+t+u=2m_W^2$.

The contributions to the $e^-e^+\to W^-W^+$ process come from s-channel
and t-channel. The s-channel amplitude is given by 
\begin{align}
{\cal M}_{sXY}^{WW} &= 
\bar{v}(p_2)\gamma_{\rho}P_YP_Xu(p_1)
V^{\mu\nu\rho}(-k_1,-k_2,k_1+k_2)
\epsilon_\mu^*(k_1)
\epsilon_\nu^*(k_2)
\nonumber\\
&\ \ \times 
\sum_{a}
g_{V_{a}ee}^{X}g_{V_{a}WW}
\frac{1}{(s-m_{V_{a}}^{2}) + i m_{V_{a}} \Gamma_{V_{a}}}
\nonumber\\
&=:J_{sXY}^{\mu\nu}\epsilon_\mu^*(k_1)\epsilon_\nu^*(k_2),
\label{Eq:Amplitude-ee-to-WW-S}
\end{align}
where 
$V^{\alpha\beta\gamma}(q,k_{-},k_{+}):=
\left[
 (q-k_{-})^{\gamma} g^{\alpha\beta}
+(k_{-}-k_{+})^{\alpha} g^{\beta\gamma}
+(k_{+}-q)^{\beta} g^{\gamma\alpha}
\right]$,
$q^\mu=(p_1+p_2)^\mu=(k_1+k_2)^\mu$,
$g_{V_{a}ff}^{L/R}$ are the left-(right-)handed couplings of $f$ to
$V_{a}$.
$g_{V_{a}WW}$ is the coupling constant of $V_a-W-W$, where 
$m_{V_{a}}$ and $\Gamma_{V_{a}}$ are mass and total decay width of
$V_{a}$. $q^\mu=(p_1+p_2)^\mu=(k_2+k_1)^\mu$.
The $J_{sXY}^{\mu\nu}$ can be written as
\begin{align}
 J_{sXY}^{\mu\nu}
&=
\left\{
\begin{array}{ll}
\displaystyle
A_{sX}^{WW}
\bar{v}(p_2) \gamma_{\rho} P_Xu(p_1)
V^{\mu\nu\rho}(-k_1,-k_2,k_1+k_2)
&\mbox{for}\ X=Y \\
\displaystyle
0
&\mbox{for}\ X\not=Y \\
\end{array}
\right.,
\label{Eq:JsXY}
\end{align}
where
\begin{align}
A_{sX}^{WW}:=
\sum_{a}
g_{V_{a}ee}^{X} g_{V_{a}WW}
\frac{1}{(s-m_{V_{a}}^{2}) + i m_{V_{a}} \Gamma_{V_{a}}}.
\label{AsX-WW}
\end{align}
The t-channel amplitude is given by
\begin{align}
{\cal M}_{tXY}^{WW} &= 
-\epsilon_\mu^*(k_1) 
\epsilon_\nu^*(k_2)
\bar{v}(p_2) P_Y
\gamma^\nu
\left\{
\sum_{n}
g_{We\nu^{(n)}}^{X}g_{We\nu^{(n)}}^{Y}
\frac{(\cancel{p}_1-\cancel{k}_1)+m_{\nu^{(n)}}}
{(t-m_{\nu^{(n)}}^{2})}
\right\}
\gamma^\mu
P_X
u(p_1)
\nonumber\\
&=:J_{tXY}^{\mu\nu}
\epsilon_\mu^*(k_1)
\epsilon_\nu^*(k_2).
\label{Eq:Amplitude-ee-to-WW-T}
\end{align}
The $J_{tXY}^{\mu\nu}$ can be written as
\begin{align}
 J_{tXY}^{\mu\nu}
&=
\left\{
\begin{array}{ll}
\displaystyle
-A_{tX}^{WW}
\bar{v}(p_2)
\gamma^\nu
(\cancel{p}_1-\cancel{k}_1)
\gamma^\mu
P_X
u(p_1)
&\mbox{for}\ X=Y \\
\displaystyle
-B_t^{WW}
\bar{v}(p_2)
\gamma^\nu
\gamma^\mu
P_X
u(p_1)
&\mbox{for}\ X\not=Y \\
\end{array}
\right.,
\allowdisplaybreaks[1]\nonumber\\
\label{Eq:JtXY}
\end{align}
where
\begin{align}
A_{tX}^{WW}:=
\sum_{n}
\frac{(g_{We\nu^{(n)}}^{X})^2}
{t-m_{\nu^{(n)}}^{2}},\ \ 
B_{t}^{WW}:=
\sum_{n}
\frac{g_{We\nu^{(n)}}^{L}g_{We\nu^{(n)}}^{R}m_{\nu^{(n)}}}
{t-m_{\nu^{(n)}}^{2}}.
\label{AtX-WW}
\end{align}

\subsubsection{Squared amplitude}

The squared amplitude of the $e^-_{X}e^+_{\bar{Y}}\to W^-W^+$ process
is given by 
\begin{align}
\left|{\cal M}_{XY}^{WW}\right|^2 
=
\left|{\cal M}_{sXY}^{WW}+ {\cal M}_{tXY}^{WW}\right|^2 
=
\left|{\cal M}_{sXY}^{WW}\right|^{2}
+ \left|{\cal M}_{tXY}^{WW}\right|^{2}
+ {\cal M}_{sXY}^{WW} {\cal M}_{tXY}^{WW\dag} 
+ {\cal M}_{sXY}^{WW\dag} {\cal M}_{tXY}^{WW}.
\label{Eq:MXY-squared}
\end{align}
The third and forth terms stand for the interference terms between 
s- and t-channel.

First, from Eq.~(\ref{Eq:Amplitude-ee-to-WW-S}),
the s-channel contribution $|{\cal M}_{sXY}^{WW}|^{2}$ is given by
\begin{align}
|{\cal M}_{sXY}^{WW}|^{2}
&= \sum_{\text{spins}}
\epsilon^{*}_{\mu}(k_1) \epsilon^{*}_{\nu}(k_2)
\epsilon_{\mu'}(k_1) \epsilon_{\nu'}(k_2)
J_{sXY}^{\mu\nu} \bar{J}_{sXY}^{\mu'\nu'}
\nonumber\\
&= 
\left( - g_{\mu\mu'} + \frac{k_{1\mu} k_{1\mu'}}{k_{1}^{2}}\right)
\left( - g_{\nu\nu'} + \frac{k_{2\nu} k_{2\nu'}}{k_{2}^{2}}\right)
\sum_{\text{spins}}
J_{sXY}^{\mu\nu} \bar{J}_{sXY}^{\mu'\nu'}.
\label{Eq:M_sXY-WW}
\end{align}
Substituting Eq.~(\ref{Eq:JsXY}) into Eq.~(\ref{Eq:M_sXY-WW}), 
we find
\begin{align}
|{\cal M}_{sXY}^{WW}|^{2}
&=
\left\{
\begin{array}{ll}
\displaystyle
4s^2|A_{sX}^{WW}|^2A(s,t,u)
&\mbox{for}\ X=Y \\
\displaystyle
0
&\mbox{for}\ X\not=Y \\
\end{array}
\right.,
\label{Eq:MsXY-WW-final}
\end{align}
where 
\begin{align}
A(s,t,u):=
\left(\frac{tu}{m_W^4}-1\right)
\left(\frac{1}{4}-\frac{m_W^2}{s}+3\frac{m_W^4}{s^2}\right)
+\frac{s}{m_W^2}-4.
\label{Eq:Astu}
\end{align}

Second, from Eq.~(\ref{Eq:Amplitude-ee-to-WW-T})
the t-channel contribution $|{\cal M}_{tXY}^{WW}|^{2}$ is given
by
\begin{align}
|{\cal M}_{tXY}^{WW}|^{2}
&= \sum_{\text{spins}}
\epsilon^{*}_{\mu}(k_1) \epsilon^{*}_{\nu}(k_2)
\epsilon_{\mu'}(k_1) \epsilon_{\nu'}(k_2)
J_{tXY}^{\mu\nu} \bar{J}_{tXY}^{\mu'\nu'}
\nonumber\\
&= 
\left( - g_{\mu\mu'} + \frac{k_{1\mu} k_{1\mu'}}{k_{1}^{2}}\right)
\left( - g_{\nu\nu'} + \frac{k_{2\nu} k_{2\nu'}}{k_{2}^{2}}\right)
\sum_{\text{spins}}
J_{tXY}^{\mu\nu} \bar{J}_{tXY}^{\mu'\nu'}.
\label{Eq:M_tXY-WW}
\end{align}
Substituting Eq.~(\ref{Eq:JtXY}) into Eq.~(\ref{Eq:M_tXY-WW}), 
we find 
\begin{align}
|{\cal M}_{tXY}^{WW}|^{2}
&=
\left\{
\begin{array}{ll}
\displaystyle
|t A_{tX}^{WW}|^{2}\cdot
4 E(s,t,u)
&\mbox{for}\ X=Y \\
\displaystyle
|tB_{t}^{WW}|^{2} 
4\biggl[ 
\frac{2m_W^2}{t^2}\left(\frac{tu}{m_W^4}-1\right)
+\frac{2s}{t^2}+\frac{s}{2m_W^4}
\biggr]
&\mbox{for}\ X\not=Y \\
\end{array}
\right.,
\label{Eq:MtXY-WW-final}
\end{align}
where 
\begin{align}
E(s,t,u):=
\left(\frac{tu}{m_W^4}-1\right)
\left(\frac{1}{4}+\frac{m_W^4}{t^2}\right)
+\frac{s}{m_W^2}.
\label{Eq:Estu}
\end{align}

Finally, we consider the interference terms in
Eq.~(\ref{Eq:MXY-squared}).
Due to ${\cal M}_{sXY}=0$ for $X\not=Y$, 
\begin{align}
{\cal M}_{sXY} {\cal M}_{tXY}^{\dag}
={\cal M}_{sXY}^\dag {\cal M}_{tXY}
=0.
\end{align}
Therefore, the non-zero values of the interference terms are given by
\begin{align}
{\cal M}_{sXX}^{WW}{\cal M}_{tXX}^{WW\dag} 
&= \sum_{\text{spins}}
\epsilon^{*}_{\mu}(k_1) \epsilon^{*}_{\nu}(k_2)
\epsilon_{\mu'}(k_1) \epsilon_{\nu'}(k_2)
J_{sXX}^{\mu\nu} \bar{J}_{tXX}^{\mu'\nu'}
\nonumber\\
&= 
\left( - g_{\mu\mu'} + \frac{k_{+\mu} k_{+\mu'}}{k_{+}^{2}}\right)
\left( - g_{\nu\nu'} + \frac{k_{-\nu} k_{-\nu'}}{k_{-}^{2}}\right)
\sum_{\text{spins}}
J_{sXX}^{\mu\nu} \bar{J}_{tXX}^{\mu'\nu'}.
\label{Eq:M_iXY-WW}
\end{align}
Substituting Eqs.~(\ref{Eq:JsXY}) and (\ref{Eq:JtXY}) into
Eq.~(\ref{Eq:M_iXY-WW}), we find
\begin{align}
{\cal M}_{sXY}^{WW}{\cal M}_{tXY}^{WW\dag}
&=
\left\{
\begin{array}{ll}
\displaystyle
A_{sX}^{WW}A_{tX}^{WW*}\cdot 4stI(s,t,u)
&\mbox{for}\ X=Y \\
\displaystyle
0
&\mbox{for}\ X\not=Y \\
\end{array}
\right.,
\label{Eq:MiXY-WW-final}
\end{align}
where
\begin{align}
I(s,t,u) &= \left(\frac{tu}{m_W^{4}} - 1 \right)
\left(\frac{1}{4} - \frac{m_W^{2}}{2s} - \frac{m_W^{4}}{st} \right)
 + \frac{s}{m_W^{2}} - 2 + 2\frac{m_W^{2}}{t}.
\label{Eq:Istu}
\end{align}

From Eqs~(\ref{Eq:MsXY-WW-final}), (\ref{Eq:MtXY-WW-final}), and
(\ref{Eq:MiXY-WW-final}), the total amplitude of
$e_X^-e_{\bar{Y}}^+\to W^-W^+$ for $X=Y$ given by 
\begin{align}
|{\cal M}^{WW}_{XY}|^{2}
&= 
4s^2|A_{sX}^{WW}|^2A(s,t,u)
+
4t^2|A_{tX}^{WW}|^{2}
E(s,t,u)
\nonumber\\
&\ \ \
+4st
\left(
A_{sX}^{WW}A_{tX}^{WW*}
+A_{sX}^{WW*}A_{tX}^{WW}
\right)
I(s,t,u),
\label{Eq:ee-to-WW-X=Y}
\end{align}
where $A(s,t,u)$, $E(s,t,u)$, and $I(s,t,u)$ are defined in
Eqs.~(\ref{Eq:Astu}), (\ref{Eq:Estu}), and (\ref{Eq:Istu}),
respectively.
The total amplitude of $e_X^-e_{\bar{Y}}^+\to W^-W^+$ for $X\not=Y$
given by 
\begin{align}
|{\cal M}^{WW}_{XY}|^{2}
&= 
|tB_{t}^{WW}|^{2} \biggl[ \frac{4s}{t^{2}} + \frac{s}{m_W^{4}} + \frac{4}{m_W^{2}}\left(\frac{u}{t} - \frac{m_W^{4}}{t^{2}}\right)
\biggr].
\label{Eq:ee-to-WW-Xnot=Y}
\end{align}

\subsubsection{Cross section}

For the $e^-_Xe^+_{\bar{Y}}\to W^-W^+$ process, the cross section of the
initial states of the polarized electron and positron is given by
\begin{align}
\frac{d\sigma^{WW}}{d\cos\theta}(P_{e^-},P_{e^+},\cos\theta)
&=
\frac{1}{4}
\bigg\{
(1-P_{e^-})(1+P_{e^+})
\frac{d\sigma_{LL}^{WW}}{d\cos\theta}
+(1+P_{e^-})(1-P_{e^+})
\frac{d\sigma_{RR}^{WW}}{d\cos\theta}
\nonumber\\
&\ \ \ \
+(1-P_{e^-})(1-P_{e^+})
\frac{d\sigma_{LR}^{WW}}{d\cos\theta}
+(1+P_{e^-})(1+P_{e^+})
\frac{d\sigma_{RL}^{WW}}{d\cos\theta}
\bigg\},
\label{Eq:dsigma-ee-to-WW}
\end{align}
where $P_{e^-}$ and $P_{e^+}$ are the initial polarizations of the
electron and positron 
\begin{align}
\frac{d\sigma_{XY}^{WW}}{d\cos\theta}(\cos\theta):=
\frac{d\sigma}{d\cos\theta}(e_{X}^-e_{\bar{Y}}^+\to W^-W^+)
=\frac{\beta_W}{32\pi s} {|{\cal M}^{WW}_{XY}|^{2}},
\label{Eq:dsigma-ee-to-WW-XY}
\end{align}
where 
${\cal M}_{XY}^{WW}$ are given in Eqs.~(\ref{Eq:ee-to-WW-X=Y}) and 
(\ref{Eq:ee-to-WW-Xnot=Y})

The total cross section of the $e^-e^+\to W^-W^+$ process with the
initial polarizations can be defined by integrating the
differential cross section in Eq.~(\ref{Eq:dsigma-ee-to-WW}) with the
angle $\theta$
\begin{align}
 \sigma^{WW}(P_{e^-},P_{e^+}):=
 \int_{-1}^{1}
 \frac{d\sigma^{WW}}{d\cos\theta}(P_{e^-},P_{e^+},\cos\theta)
 d\cos\theta,
\end{align}
where the minimum and maximal values of $\cos\theta$ are determined
by each detector and we cannot use date near $\cos\theta\simeq \pm1$.
The total cross section of $e^-e^+\to W^-W^+$ is given by
\begin{align}
\sigma^{WW}(P_{e^-},P_{e^+})
&=\frac{1}{4}
(1-P_{e^-})(1+P_{e^+})
\sigma_{LL}^{WW}
+\frac{1}{4}
(1+P_{e^-})(1-P_{e^+})
\sigma_{RR}^{WW}
\nonumber\\
&\ \ \
+\frac{1}{4}
(1-P_{e^-})(1-P_{e^+})
\sigma_{LR}^{WW}
+\frac{1}{4}
(1+P_{e^-})(1+P_{e^+})
\sigma_{RL}^{WW},
\label{Eq:sigma-total-WW}
\end{align}
where 
\begin{align}
\sigma_{XY}^{WW}&:=
 \int_{-1}^{1}
 \frac{d\sigma_{XY}^{WW}}{d\cos\theta}(\cos\theta)
 d\cos\theta,
\label{Eq:sigma-total-WW-XY}
\end{align}
where $X=L,R$; $d\sigma_{XY}^{WW}/d\cos\theta$ 
are given in Eq.~(\ref{Eq:dsigma-ee-to-WW-XY}).

The statistical error of the cross section 
$\sigma^{WW}(P_{e^-},P_{e^+})$ is given by
\begin{align}
\Delta \sigma^{WW}(P_{e^-},P_{e^+})=
 \frac{\sigma^{WW}(P_{e^-},P_{e^+})}
 {\sqrt{N^{WW}}},\ \ 
 N^{WW} =  L_{\rm int} \cdot
 \sigma^{WW}(P_{e^-},P_{e^+}),
\label{Eq:stat-error-sigma-WW}
\end{align}
where $L_{\rm int}$ is an integrated luminosity, and 
$N^{WW}$ is the number of events for the $e^-e^+\to W^-W^+$ process.
Note that the $W$ boson cannot be observed directly, so
we need to choose the decay modes of the $W$ boson, and then the
available number of events must be $N^{WW}$ multiplied by the branching
ratio of each selected decay mode. The amount of the deviation of the
cross section of the $e^-e^+\to W^-W^+$ process from the SM in the GHU
model is given by
\begin{align}
\Delta_\sigma^{WW}(P_{e^-},P_{e^+}):=
\frac{
\left[\sigma^{WW}(P_{e^-},P_{e^+})\right]_{\rm GHU}}
{\left[\sigma^{WW}(P_{e^-},P_{e^+})\right]_{\rm SM}}-1,
\label{Eq:Delta_sigma-WW}
\end{align}
where $\left[\sigma^{WW}(P_{e^-},P_{e^+})\right]_{\rm GHU}$ 
and $\left[\sigma^{WW}(P_{e^-},P_{e^+})\right]_{\rm SM}$ 
stand for the cross sections of the $e^-e^+\to W^-W^+$ process in
the SM and the GHU model, respectively.
The same notation is used for other cases in the followings.

\subsubsection{Left-right asymmetry}

We define an observable left-right asymmetry 
\cite{MoortgatPick:2005cw,Abe:1994wx,Abe:1996nj,Funatsu:2020haj,Funatsu:2022spb} 
of the $e^-e^+\to W^-W^+$ process as
\begin{align}
A_{LR}^{WW}(P_{e^-},P_{e^+})
:=\frac{\sigma^{WW}(P_{e^-},P_{e^+})-\sigma^{WW}(-P_{e^-},-P_{e^+})}
{\sigma^{WW}(P_{e^-},P_{e^+})+\sigma^{WW}(-P_{e^-},-P_{e^+})}
\label{Eq:ALR-WW-def}
\end{align}
for $P_{e^-}<0$ and $|P_{e^-}|>|P_{e^+}|$.

The statistical error of the left-right asymmetry is given by
\begin{align}
 \Delta A_{LR}^{WW}(P_{e^-},P_{e^+}) &=
 2\frac{\sqrt{N_{L}^{WW}N_{R}^{WW}}
 \left(\sqrt{N_{L}^{WW}}+\sqrt{N_{R}^{WW}}\right)}
 {(N_{L}^{WW}+N_{R}^{WW})^2},
\label{Eq:Error-A_LR-WW}
\end{align}
where $N_{L}^{WW}=L_{\rm int} \, \sigma^{WW}(P_{e^-},P_{e^+})$
and $N_{R}^{WW}=L_{\rm int} \, \sigma^{WW}(-P_{e^-},-P_{e^+})$
are the numbers of the events 
for $P_{e^-}<0$ and $|P_{e^-}|>|P_{e^+}|$.
The amount of the deviation from the SM in Eq.~(\ref{Eq:ALR-WW-def})
is characterized by
\begin{align}
\Delta_{A_{LR}}^{WW}(P_{e^-},P_{e^+}) &:=
 \frac{\left[A_{LR}^{WW}(P_{e^-},P_{e^+})\right]_{\rm GHU}}
{\left[A_{LR}^{WW}(P_{e^-},P_{e^+})\right]_{\rm SM}}-1.
\label{Eq:Delta_A_LR-WW}
\end{align}

\subsubsection{Asymptotic behavior}

We consider the asymptotic behavior of $e^-e^+\to W^-W^+$ for large $s$
to confirm that the Goldstone boson equivalence theorem
\cite{Cornwall:1974km,Chanowitz:1985hj,Peskin:1995ev}
is satisfied in the GHU model.
From Eqs.~(\ref{Eq:dsigma-ee-to-WW}), (\ref{Eq:dsigma-ee-to-WW-XY}), and
(\ref{Eq:sigma-total-WW}), very large cross section appears
if the total squared amplitude of $|{\cal M}_{XY}^{WW}|^2$ contain
$O(s^2)$ terms.
The $O(s)$ terms of the squared amplitude do not break unitality 
because the Froissart bound  \cite{Froissart:1961ux,Langacker:2017uah}
is satisfied.
In this section we consider the $O(s^2)$ and $O(s)$ terms of
the squared amplitude. In the SM, the $O(s)$ terms vanish.

The total amplitudes of $e_X^-e_{\bar{Y}}^+\to W^-W^+$ are given in
Eqs~(\ref{Eq:ee-to-WW-X=Y}) and (\ref{Eq:ee-to-WW-Xnot=Y}).
First, from Eqs.~(\ref{AsX-WW}) and (\ref{AtX-WW}), 
the asymptotic behavior of 
$A_{sX}^{WW}$, $A_{tX}^{WW}$, and $B_{t}^{WW}$ for 
large $s$ is given by
\begin{align}
sA_{sX}^{WW}&\sim
\sum_{a}
g_{V_{a}ee}^{X} g_{V_{a}WW}
\left(1+\frac{m_{V_a}^2}{s}\right),
\nonumber\\
tA_{tX}^{WW}&\sim
\sum_{n}
(g_{We\nu^{(n)}}^{X})^2
\left(1+\frac{m_{\nu^{(n)}}^2}{t}\right),
\nonumber\\
tB_{t}^{WW}&\sim
\sum_{n}
g_{We\nu^{(n)}}^{L}g_{We\nu^{(n)}}^{R}m_{\nu^{(n)}}
\left(1+\frac{m_{\nu^{(n)}}^2}{t}\right),
\label{Eq:Asymptotic-AB}
\end{align}
where we ignore the decay width $\Gamma_{V_a}$ for simplicity.
Second, from Eqs.~(\ref{Eq:Astu}), (\ref{Eq:Estu}), and (\ref{Eq:Istu}),
the asymptotic behavior of $A(s,t,u)$, $E(s,t,u)$, and $I(s,t,u)$ for
large $s$ is given by
\begin{align}
4A(s,t,u)&\sim
\frac{tu}{m_W^4}
-\frac{4tu}{sm_W^2}
+\frac{4s}{m_W^2}
\nonumber\\
4E(s,t,u)&\sim
\frac{tu}{m_W^4}
+\frac{4s}{m_W^2},
\nonumber\\
4I(s,t,u) &\sim
\frac{tu}{m_W^{4}}
-\frac{2tu}{sm_W^{2}}
+ \frac{4s}{m_W^{2}}.
\label{Eq:Asymptotic-AEI}
\end{align}
Substituting Eqs.~(\ref{Eq:Asymptotic-AB}) and (\ref{Eq:Asymptotic-AEI})
into Eq.~(\ref{Eq:ee-to-WW-X=Y}), we find the asymptotic behavior of 
$|{\cal M}^{WW}_{XY}|^{2}$ for $X=Y$:
\begin{align}
|{\cal M}^{WW}_{XY}|^{2}
&\sim
\left(
\sum_{a}
g_{V_{a}ee}^{X} g_{V_{a}WW}
+
\sum_{n}
(g_{We\nu^{(n)}}^{X})^2
\right)
\nonumber\\
&\ \ \
\times
\bigg[
\left(
\sum_{a}
g_{V_{a}ee}^{X} g_{V_{a}WW}
+
\sum_{n}
(g_{We\nu^{(n)}}^{X})^2
\right)
\left(\frac{tu}{m_W^4}+\frac{4s}{m_W^2}\right)
\nonumber\\
&\ \ \ \ \
-2
\left\{
\sum_{b}
g_{V_{b}ee}^{X} g_{V_{b}WW}
-
\sum_{b}
g_{V_{b}ee}^{X} g_{V_{b}WW}\frac{m_{V_b}^2}{m_W^2}
\right\}
\frac{tu}{sm_W^2}
\nonumber\\
&\ \ \ \ \
+2
\left(
\sum_{m}
(g_{We\nu^{(m)}}^{X})^2
\frac{m_{\nu^{(m)}}^2}{m_W^2}
\right)
\frac{u}{m_W^2}
\bigg].
\label{Eq:Asymptotic-M-WW-X=Y}
\end{align}
Substituting Eqs.~(\ref{Eq:Asymptotic-AB}) and (\ref{Eq:Asymptotic-AEI})
into Eq.~(\ref{Eq:ee-to-WW-Xnot=Y}), we find the asymptotic behavior of 
$|{\cal M}^{WW}_{XY}|^{2}$ for $X\not=Y$:
\begin{align}
|{\cal M}^{WW}_{XY}|^{2}
\sim
\left(\sum_ng_{We\nu^{(n)}}^{L}g_{We\nu^{(n)}}^{R}
\frac{m_{\nu^{(n)}}}{m_W}\right)^2
\frac{s}{m_W^{2}}.
\label{Eq:Asymptotic-M-WW-Xnot=Y}
\end{align}

Here we summarize the above results. From the asymptotic behavior 
of $|{\cal M}^{WW}_{XY}|^{2}$ for $X=Y$
in Eq.~(\ref{Eq:Asymptotic-M-WW-X=Y}), $O(s^2)$ terms disappear when the
following condition is satisfied:
\begin{align}
\sum_{a}
g_{V_{a}ee}^{X} g_{V_{a}WW}
+
\sum_{n}
(g_{We\nu^{(n)}}^{X})^2
=0.
\label{Eq:Unitality-condition-1}
\end{align}
The $O(s)$ terms also disappear if this condition is satisfied.
Next, from the asymptotic behavior of $|{\cal M}^{WW}_{XY}|^{2}$ for
$X\not=Y$ in Eq.~(\ref{Eq:Asymptotic-M-WW-Xnot=Y}), $O(s)$ terms
disappear when the following condition is satisfied:
\begin{align}
\sum_ng_{We\nu^{(n)}}^{L}g_{We\nu^{(n)}}^{R}\frac{m_{\nu^{(n)}}}{m_W}=0.
\label{Eq:Unitality-condition-2}
\end{align}

\begin{table}[tbh]
{
\begin{center}
\begin{tabular}{|c|c|c|c|c|c|}
\hline
\rowcolor[gray]{0.9}
 Coupling&$n=0$&$n=1$&$n=2$&$n=3$&$n=4$
\\\hline 
 $g_{\gamma^{(n)}WW}$
 &$+\sin\theta_W^0$&$-0.00029313$&$-0.00001057$&$-0.00000185$&$+0.00000052$\\
 $g_{Z^{(n)}WW}$
 &$+0.87715447$&$-0.00018818$&$+0.00000002$&$-0.00000679$&$-0.00000002$\\
 $g_{Z_R^{(n)}WW}$
 &$-$&$+0.00027271$&$+0.00000983$&$+0.00000163$&$+0.00000047$\\
 \hline
 $g_{\gamma^{(n)}ee}^{L}$
 &$-\sin\theta_W^0$&$-2.75873260$&$-1.08511541$&$-0.39362141$&$-0.23177025$\\
 $g_{\gamma^{(n)}ee}^{R}$
 &$-\sin\theta_W^0$&$+0.10708192$&$-0.07393235$&$+0.06029203$&$-0.05236728$\\
 \hline
 $g_{Z^{(n)}ee}^{L}$
 &$-0.30576326$&$-1.76206116$&$-0.00643559$&$-0.69313977$&$-0.00209004$\\
 $g_{Z^{(n)}ee}^{R}$
 &$+0.26291828$&$-0.05843186$&$-0.00396529$&$+0.04034314$&$+0.00296713$\\
 \hline
 $g_{Z_R^{(n)}ee}^{L}$
 &$-$&$-1.04437969$&$-0.41581800$&$-0.14941965$&$-0.08774238$\\
 $g_{Z_R^{(n)}ee}^{R}$
 &$-$&$0$&$0$&$0$&$0$\\
 \hline
 $g_{We\nu_e^{(n)}}^{L}$
 &$+0.99764683$ &$-0.01669268$ &$-0.00001276$ &$+0.00209568$&$-0.00000029$\\
 $g_{We\nu_e^{(n)}}^{R}$
 &$0$&$0$&$0$&$0$&$0$\\
\hline
 $g_{We\nu_{e2}^{(n)}}^{L}$
 &$-$
 &$-0.01669268$&$-0.00001276$&$+0.00209568$&$-0.00000029$\\
 $g_{We\nu_{e2}^{(n)}}^{R}$
 &$-$
 &$0$&$0$&$0$&$0$\\
\hline
\end{tabular}
 \caption{\small
 Coupling constants of $W$ bosons to neutral gauge bosons,
 electrons to neutral gauge bosons,
 $W$ bosons to electron and zero mode and KK neutrinos are shown
 up to $n=4$
 in the GHU model with the parameter set of $m_{\rm KK}=13$\,TeV and 
 $\theta_H=0.10$, and negative bulk masses of the leptons $(A_-)$.
 The values of the coupling constants for neutral gauge bosons to $W$
 bosons and leptons are given in units of  $g_w$.
 The values of the coupling constants for $W$ boson to leptons are in
 units of $g_w/\sqrt{2}$.
 $g_w=e/\sin\theta_W^0$, $\sin^2\theta_W^0= 0.2306$ (input).
 In the SM $\sin^2\theta_W(\overline{\mbox{MS}})=0.23122\pm 0.00004$
 \cite{ParticleDataGroup:2022pth}.
 When the value is less than $10^{-8}$, we write $0$.
 When there is no corresponding coupling constant, we write the symbol
 $-$ in that field. 
 }
\label{Table:KK-mode-couplings}
\end{center}
}
\end{table}

We check that the unitality is preserved in the SM and the GHU below
by using Eqs.~(\ref{Eq:Unitality-condition-1}) and 
(\ref{Eq:Unitality-condition-2}).
First, in the SM, $V_a=\gamma,Z$ and
$\nu^{(n)}=\nu_e$, relevant coupling  constants are given by 
\begin{align}
&g_{\gamma ee}^{L}=g_{\gamma ee}^{R}=-e,\ \ 
g_{\gamma WW}=e,\ \ 
g_{ZWW}=g_w\cos\theta_W,\ \ 
g_{We\nu_e}^{L}=\frac{g_w}{\sqrt{2}},\ \ 
g_{We\nu_e}^{R}=0,
\nonumber\\
&g_{Zee}^{L}=
\frac{g_w}{\cos\theta_W}\left(-\frac{1}{2}+\sin^2\theta_W\right),
\ \ 
g_{Zee}^{R}=\frac{g_w}{\cos\theta_W}\sin^2\theta_W,
\end{align}
where $g_w=e/\sin\theta_W$. Therefore, we find
\begin{align}
&\sum_{a}g_{V_{a}ee}^{L}g_{V_{a}WW}
= g_{\gamma ee}^{L}g_{\gamma WW}+g_{Ze}^{L}g_{ZWW}
=-\frac{g_w^2}{2},
\ \ \
\sum_{n}(g_{We\nu^{(n)}}^{L})^2
= (g_{We\nu}^{L})^2
=\frac{g_w^2}{2},
\nonumber\\
&\sum_{a}
g_{V_{a}ee}^{R}g_{V_{a}WW}
= g_{\gamma ee}^{R}g_{\gamma WW}+g_{Ze}^{R}g_{ZWW}
=0,
\ \ \
\sum_{n}(g_{We\nu^{(n)}}^{R})^2
= (g_{We\nu}^{R})^2=0.
\end{align}
Substituting the above SM gauge coupling constants into
Eq.~(\ref{Eq:Unitality-condition-1}), the coefficients of the $O(s^2)$
and $O(s)$ terms of ${\cal M}_{LL}^{WW}$ and ${\cal M}_{RR}^{WW}$ 
vanish.
Next, even when we take into account the non-zero neutrino mass
$m_{\nu}\not=0$, since $g_{We\nu}^{R}=0$ in the SM, we find
\begin{align}
\sum_ng_{We\nu^{(n)}}^{L}g_{We\nu^{(n)}}^{R}\frac{m_{\nu^{(n)}}}{m_W}
=g_{We\nu}^{L}g_{We\nu}^{R}\frac{m_{\nu}}{m_W}=0.
\end{align}
Therefore, the coefficients of the $O(s)$ terms of ${\cal M}_{LR}^{WW}$
and ${\cal M}_{RL}^{WW}$ vanish.

\begin{table}[htb]
{
\begin{center}
\begin{tabular}{|c|c|c|c|c|c|c|}
\hline
\rowcolor[gray]{0.9}
 $(m_{\rm KK},\theta_H)$&SM&($13$\,TeV,\,0.10)&($25$\,TeV,\,0.05)&($50$\,TeV,\,0.025)&($100$\,TeV,\,0.0125)\\
\hline 
$\displaystyle
\sum_{a}g_{V_{a}ee}^{L}g_{V_{a}WW}$
&$-0.5$&$-0.49793312$&$-0.499484367$&$-0.49987087$&$-0.49996771$
\\
$\displaystyle
\sum_{n}(g_{We\nu^{(n)}}^{L})^2$
&$+0.5$&$+0.49793310$&$+0.499448367$&$+0.49987087$&$+0.49996771$
\\
\hline\hline
$\displaystyle
\sum_{a}g_{V_{a}ee}^{R}g_{V_{a}WW}$
&$0$&$0$ &$0$&$0$&$0$
\\
$\displaystyle
\sum_{n}(g_{We\nu^{(n)}}^{R})^2$
&$0$&$0$ &$0$&$0$&$0$
\\
\hline
\hline
$\displaystyle
\sum_{n}g_{We\nu^{(n)}}^{L}g_{We\nu^{(n)}}^{R}
\frac{m_{\nu^{(n)}}}{m_W}$
&$0$&$0$&$0$&$0$&$0$
\\
\hline
\end{tabular}
 \caption{\small
The values of the coupling summation are summarized for several sets of
 parameters  
$(m_{\rm KK},\theta_H)$ with negative bulk masses of the leptons in the
 units of $g_w^2$. 
 When the value is less than $10^{-8}$, we write $0$.
 }
\label{Table:Coupling-summation}
\end{center}
}
\end{table}

We consider the GHU model with the parameter set for
$m_{\rm KK}=13$\,TeV, $\theta_H=0.10$, and negative  bulk masses of the
leptons. For the GHU model, 
$V_a=\gamma,Z,\gamma^{(n)},Z^{(n)},Z_R^{(m)}$ $(n=1,2,...)$ and
$\nu^{(n)}=\nu_e,\nu_e^{(n)}$ $(n=1,2,...)$, relevant coupling 
constants are given in Table~\ref{Table:KK-mode-couplings},
where the following calculation of the sum of coupling constants adds up
to a sufficiently large KK mode.
We find
\begin{align}
\sum_{a}g_{V_{a}ee}^{L}g_{V_{a}WW}
&\simeq -0.49793312g_w^2,
\ \ \
\sum_{n}(g_{We\nu^{(n)}}^{L})^2
\simeq +0.49793310g_w^2,
\allowdisplaybreaks[1]\nonumber\\
\left|
\sum_{a}
g_{V_{a}ee}^{R}g_{V_{a}WW}\right|
&
< 10^{-8}\times g_w^2,
\ \ \
\left|
\sum_{n}(g_{We\nu^{(n)}}^{R})^2\right|
< 10^{-8}\times g_w^2.
\end{align}
We numerically find 
\begin{align}
&\sum_{a}g_{V_{a}ee}^{L}g_{V_{a}WW}
+\sum_{n}(g_{We\nu^{(n)}}^{L})^2
\simeq -2\times 10^{-8}g_w^2,
\nonumber\\
&\left|
\sum_{a}g_{V_{a}ee}^{R}g_{V_{a}WW}
+\sum_{n}(g_{We\nu^{(n)}}^{R})^2\right|
< 10^{-8}\times g_w^2.
\end{align}
Therefore, the $O(s^2)$ and $O(s)$ terms of 
${\cal M}_{LL}^{WW}$ and ${\cal M}_{RR}^{WW}$
are well suppressed.
We also find
\begin{align}
\left|
\sum_ng_{We\nu^{(n)}}^{L}g_{We\nu^{(n)}}^{R}\frac{m_{\nu^{(n)}}}{m_W}
\right|
< 10^{-8}\times g_w^2.
\end{align}
The $O(s)$ term of ${\cal M}_{LR}^{WW}$ and ${\cal M}_{RL}^{WW}$
is sufficiently suppressed.
Several other parameter sets are also summarized in
Table~\ref{Table:Coupling-summation}.

\begin{table}[tbh]
{
\begin{center}
\begin{tabular}{|c|c|c|c|c|c|}
\hline
\rowcolor[gray]{0.9}
 Coupling&$n=0$&$n=1$&$n=2$&$n=3$&$n=4$
\\\hline 
 $g_{\gamma^{(n)}WW}$
 &$+\sin\theta_W^0$&$-0.00029370$&$-0.00001060$&$-0.00000185$&$-0.00000052$\\
 $g_{Z^{(n)}WW}$
 &$+0.87647017$&$-0.00018736$&$+0.00000002$&$-0.00000676$&$-0.00000002$\\
 $g_{Z_R^{(n)}WW}$
 &$-$&$+0.00027211$&$+0.00000980$&$+0.00000162$&$+0.00000047$\\
 \hline
 $g_{\gamma^{(n)}ee}^{L}$
 &$-\sin\theta_W^0$&$+0.10727367$&$-0.07406031$&$+0.06039443$&$-0.05245504$\\
 $g_{\gamma^{(n)}ee}^{R}$
 &$-\sin\theta_W^0$&$-2.76765185$&$-1.08788152$&$-0.39449879$&$-0.23232603$\\
 \hline
 $g_{Z^{(n)}ee}^{L}$
 &$-0.30602661$&$+0.06795725$&$+0.00461504$&$-0.04691696$&$-0.00345321$\\
 $g_{Z^{(n)}ee}^{R}$
 &$+0.26309454$&$+1.51736122$&$+0.00553476$&$+0.59638708$&$+0.00179601$\\
 \hline
 $g_{Z_R^{(n)}ee}^{L}$
 &$-$&$0$&$0$&$0$&$0$\\
 $g_{Z_R^{(n)}ee}^{R}$
 &$-$&$+1.38256665$&$+0.55008507$&$+0.19760182$&$+0.11605766$\\
 \hline
 $g_{We\nu_e^{(n)}}^{L}$
 &$+0.998816639$&$-0.03534296$&$0$&$0$&$0$\\
 $g_{We\nu_e^{(n)}}^{R}$
 &$0$&$-0.00176655$&$-0.01669757$&$-0.00000003$&$+0.002095116$\\
\hline
 $g_{We\nu_{e2}^{(n)}}^{L}$
 &$-$&$+0.03534296$&$0$&$0$&$0$\\
 $g_{We\nu_{e2}^{(n)}}^{R}$
 &$-$&$+0.00176655$&$-0.01669757$&$-0.00000003$&$+0.002095116$\\
\hline
\end{tabular}
 \caption{\small
 Coupling constants of $W$ bosons to neutral gauge bosons,
 electrons to neutral gauge bosons,
 $W$ bosons to electron and zero mode and KK neutrinos are shown
 up to $n=4$
 in the GHU model with the parameter set of $m_{\rm KK}=13$\,TeV and 
 $\theta_H=0.10$, and positive bulk masses of the leptons $(A_+)$.
 The values of the coupling constants for neutral gauge bosons to $W$
 bosons and leptons are given in units of  $g_w$.
 The values of the coupling constants for $W$ boson to leptons are in
 units of $g_w/\sqrt{2}$.
 $g_w=e/\sin\theta_W^0$, $\sin^2\theta_W^0= 0.2306$ (input).
 In the SM $\sin^2\theta_W(\overline{\mbox{MS}})=0.23122\pm 0.00004$
 \cite{ParticleDataGroup:2022pth}.
 When the value is less than $10^{-8}$, we write $0$.
 When there is no corresponding coupling constant, we write the symbol
 $-$ in that field. 
 }
\label{Table:KK-mode-couplings-positive}
\end{center}
}
\end{table}

\begin{table}[htb]
{
\begin{center}
\begin{tabular}{|c|c|c|c|c|c|c|}
\hline
\rowcolor[gray]{0.9}
 $(m_{\rm KK},\theta_H)$&SM&($13$\,TeV,\,0.10)&($25$\,TeV,\,0.05)&($50$\,TeV,\,0.025)&($100$\,TeV,\,0.0125)\\
\hline 
$\displaystyle
\sum_{a}g_{V_{a}ee}^{L}g_{V_{a}WW}$
&$-0.5$&$-0.50006646$&$-0.50001791$&$-0.50000448$&$-0.50000112$
\\
$\displaystyle
\sum_{n}(g_{We\nu^{(n)}}^{L})^2$
&$+0.5$&$+0.50006646$&$+0.50001791$&$+0.50000448$&$+0.50000112$
\\
\hline\hline
$\displaystyle
\sum_{a}g_{V_{a}ee}^{R}g_{V_{a}WW}$
&$0$&$-0.00028680$&$-0.00006933$&$-0.00001731$&$-0.00000433$
\\
$\displaystyle
\sum_{n}(g_{We\nu^{(n)}}^{R})^2$
&$0$&$+0.00028680$&$+0.00006933$&$+0.00001731$&$+0.00000433$
\\
\hline
\hline
$\displaystyle
\sum_{n}g_{We\nu^{(n)}}^{L}g_{We\nu^{(n)}}^{R}
\frac{m_{\nu^{(n)}}}{m_W}$
&$0$&$+0.00000003$&$+0.00000002$&$+0.00000001$&$0$
\\
\hline
\end{tabular}
 \caption{\small
The values of the coupling summation are summarized for several sets of
 parameters  
$(m_{\rm KK},\theta_H)$ with positive bulk masses of the leptons in the units of $g_w^2$.
When the value is less than $10^{-8}$, we write $0$.
 }
\label{Table:Coupling-summation-positive}
\end{center}
}
\end{table}

Next, we consider the GHU model with the parameter set for
$m_{\rm KK}=13$\,TeV, $\theta_H=0.10$, and positive  bulk masses of
leptons. For the GHU model, 
$V_a=\gamma,Z,\gamma^{(n)},Z^{(n)},Z_R^{(m)}$ $(n=1,2,...)$ and
$\nu^{(n)}=\nu_e,\nu_e^{(n)}$ $(n=1,2,...)$, relevant coupling 
constants are given in Table~\ref{Table:KK-mode-couplings-positive},
where the following calculation of the sum of coupling constants adds up
to a sufficiently large KK mode.
We find
\begin{align}
\sum_{a}g_{V_{a}ee}^{L}g_{V_{a}WW}
&\simeq -0.50006646g_w^2,
\ \ \
\sum_{n}(g_{We\nu^{(n)}}^{L})^2
\simeq +0.50006646g_w^2,
\allowdisplaybreaks[1]\nonumber\\
\sum_{a}
g_{V_{a}ee}^{R}g_{V_{a}WW}
&\simeq -0.00028680 g_w^2,
\ \ \
\sum_{n}(g_{We\nu^{(n)}}^{R})^2
 \simeq +0.00028680 g_w^2.
\end{align}
We numerically find 
\begin{align}
&\left|
\sum_{a}g_{V_{a}ee}^{L}g_{V_{a}WW}
+\sum_{n}(g_{We\nu^{(n)}}^{L})^2
\right|
< 10^{-8}\times g_w^2,
\nonumber\\
&\left|
\sum_{a}g_{V_{a}ee}^{R}g_{V_{a}WW}
+\sum_{n}(g_{We\nu^{(n)}}^{R})^2\right|
< 10^{-8}\times g_w^2.
\end{align}
Therefore, 
the $O(s^2)$ and $O(s)$ terms of 
${\cal M}_{LL}^{WW}$ and ${\cal M}_{RR}^{WW}$
are sufficiently suppressed.
We also find that
\begin{align}
\sum_ng_{We\nu^{(n)}}^{L}g_{We\nu^{(n)}}^{R}\frac{m_{\nu^{(n)}}}{m_W}
\simeq 3\times 10^{-8}g_w^2.
\end{align}
The $O(s)$ term of ${\cal M}_{LR}^{WW}$ and ${\cal M}_{RL}^{WW}$
is well suppressed.
Several other parameter sets are also summarized in
Table~\ref{Table:Coupling-summation-positive}.

\subsection{$e^-e^+\to ZZ$ process}

\subsubsection{Amplitude}

Here we consider the following t- and u-channel processes:
\begin{eqnarray}
e^-_{X}(p_1) e^+_{\bar{Y}}(p_2) \to Z_\mu(k_1) Z_\nu(k_2),
\end{eqnarray}
where $X,Y=L,R$, $\bar{L}=R$, $\bar{R}=L$;
$p_1$ and $p_2$ are the momenta of the initial states of electron
and positron; $k_1$ and $k_2$ are the momenta of the final states of $Z$
bosons.
For massive bosons $Z$ and massless electrons $e^{\pm}$,
in the center-of-mass frame, we use the following basis:
\begin{align}
p_1 &= \frac{\sqrt{s}}{2}(1,0,0,+1),
\ \ \
p_2 = \frac{\sqrt{s}}{2}(1,0,0,-1),
\nonumber\\
k_1 &= \frac{\sqrt{s}}{2}(1,0,+\beta_Z\sin\theta, +\beta_Z\cos\theta),
\ \ \
k_2 =\frac{\sqrt{s}}{2}(1,0,-\beta_Z \sin\theta, -\beta_Z\cos\theta),
\end{align}
where $\beta_Z:=\sqrt{1-\frac{4m_Z^2}{s}}$.
The Mandelstam variables are defined as
\begin{align}
&s:=(p_1+p_2) ^2=(k_1+k_2)^2,
\nonumber\\
&t:=(p_1-k_1)^2=(p_2-k_2)^2
=m_Z^2-\frac{s}{2}(1-\beta_Z\cos\theta),
\nonumber\\
&u:=(p_1-k_2)^2=(p_2-k_1)^2
=m_Z^2-\frac{s}{2}(1+\beta_Z\cos\theta),
\end{align}
where $s+t+u=2m_Z^2$.

The contributions to the $e^-e^+\to ZZ$ process come from t-channel
and u-channel. The t-channel amplitude is given by
\begin{align}
{\cal M}_{tXY}^{ZZ}
&=
 \bar{v}(p_2) \gamma^\nu
\left[
\sum_{i}
g_{Yi} P_Y
\frac{(\cancel{p}_1 - \cancel{k}_1) + m_{i}}{(p_1 - k_1)^2 - m_{i}^{2}}
\gamma^\mu
g_{Xi} P_X
\right]
u(p_1)
\epsilon_\mu^*(k_1) \epsilon_\nu^*(k_2)
\nonumber\\
&=: K_{tXY}^{\mu\nu} \epsilon_\mu^*(k_1) \epsilon_\nu^*(k_2),
\end{align}
where $P_{L/R}=(1\mp\gamma_5)/2$, and
$i$ stands for electron and KK charged leptons.
The $K_{tXY}^{\mu\nu}$ $(X,Y=L,R)$ can
be written as 
\begin{align}
K^{\mu\nu}_{tXY}
&=
\left\{
\begin{array}{ll}
\displaystyle
A_{tX}^{ZZ}
\bar{v}(p_{2}) \gamma^{\nu}
(\cancel{p}_{1} - \cancel{k}_{1})
\gamma^{\mu} P_X u(p_{1}),
&\mbox{for}\ X=Y \\
\displaystyle
B_{t}^{ZZ}
\bar{v}(p_{2}) \gamma^{\nu}
\gamma^{\mu} P_Xu(p_{1}),
&\mbox{for}\ X\not=Y \\
\end{array}
\right.,
\label{Eq:KtXY}
\end{align}
where 
\begin{align}
A_{tX}^{ZZ} :=
\sum_{i} \frac{g_{Xi}^{2}}{(p_{1} - k_{1})^{2} - m_{i}^{2}},\ \ \
B_{t}^{ZZ} :=
\sum_{i} \frac{g_{Li} g_{Ri} m_{i}}{(p_{1} - k_{1})^{2} - m_{i}^{2}}.
\end{align}
The u-channel amplitude is given by
\begin{align}
{\cal M}_{uXY}^{ZZ}&=
 \bar{v}(p_2) \gamma^\mu
\left[\sum_{i} 
g_{Yi} P_Y
\frac{(\cancel{p}_1 - \cancel{k}_2) + m_{i}}{(p_1 - k_2)^2 - m_{i}^{2}}
\gamma^\nu
g_{X} P_X 
\right]
u(p_{1})
\epsilon_\mu^*(k_1) \epsilon_\nu^*(k_2)
\nonumber\\
&=: K_{uXY}^{\mu\nu} \epsilon_\mu^*(k_1) \epsilon_\nu^*(k_2),
\end{align}
where $P_{L/R}=(1\mp\gamma_5)/2$. 
The $K_{uXY}^{\mu\nu}$ $(X,Y=L,R)$ can
be written as 
\begin{align}
K^{\mu\nu}_{uXY}
&=
\left\{
\begin{array}{ll}
\displaystyle
A_{uX}^{ZZ}
\bar{v}(p_{2})\gamma^{\mu} 
(\cancel{p}_{1} - \cancel{k}_{2})
\gamma^{\nu} P_{X}u(p_{1}),
&\mbox{for}\ X=Y \\
\displaystyle
B_{u}^{ZZ}
\bar{v}(p_{2})\gamma^{\mu}
\gamma^{\nu} P_Xu(p_{1}),
&\mbox{for}\ X\not=Y \\
\end{array}
\right.,
\label{Eq:KuXY}
\end{align}
where 
\begin{align}
A_{uX}^{ZZ}
:=\sum_{i} \frac{g_{Xi}^{2}}{(p_{1} - k_{2})^{2} - m_{i}^{2}},
\ \ \
B_{u}^{ZZ}
:=\sum_{i}\frac{g_{Li}g_{Ri} m_{i}}{(p_{1} - k_{2})^{2} - m_{i}^{2}}.
\end{align}

\subsubsection{Squared amplitude}

The squared amplitude of the $e^-_{X}e^+_{\bar{Y}}\to ZZ$ process is
given by 
\begin{align}
\left|{\cal M}_{XY}^{ZZ}\right|^2 
=
\left|{\cal M}_{tXY}^{ZZ}+ {\cal M}_{uXY}^{ZZ}\right|^2 
=
\left|{\cal M}_{tXY}^{ZZ}\right|^{2}
+ \left|{\cal M}_{uXY}^{ZZ}\right|^{2}
+ {\cal M}_{tXY}^{ZZ} {\cal M}_{uXY}^{ZZ\dag} 
+ {\cal M}_{tXY}^{ZZ\dag} {\cal M}_{uXY}^{ZZ}.
\end{align}
The third and forth terms stand for the interference terms between 
t- and u-channel.

First, the t-channel contribution $|{\cal M}_{tXY}^{ZZ}|^{2}$ is given by
\begin{align}
|{\cal M}_{tXY}^{ZZ}|^{2}
&= \sum_{\text{spins}}
\epsilon^{*}_{\mu}(k_{1}) \epsilon^{*}_{\nu}(k_{2})
\epsilon_{\mu'}(k_{1}) \epsilon_{\nu'}(k_{2})
K_{tXY}^{\mu\nu} \bar{K}_{tXY}^{\mu'\nu'}
\nonumber\\
&= 
\left( - g_{\mu\mu'} + \frac{k_{1\mu} k_{1\mu'}}{k_{1}^{2}}\right)
\left( - g_{\nu\nu'} + \frac{k_{2\nu} k_{2\nu'}}{k_{2}^{2}}\right)
\sum_{\text{spins}}
K_{tXY}^{\mu\nu} \bar{K}_{tXY}^{\mu'\nu'}.
\label{Eq:MtXY-ZZ}
\end{align}
Substituting Eq.~(\ref{Eq:KtXY}) into Eq.~(\ref{Eq:MtXY-ZZ}),
we find 
\begin{align}
|{\cal M}_{tXY}^{ZZ}|^{2}
&=
\left\{
\begin{array}{ll}
\displaystyle
|t A_{tX}^{ZZ}|^{2}
\biggl[
4 \left(\frac{u}{t} - \frac{m_Z^{4}}{t^{2}}\right) + \frac{1}{m_Z^{4}} (ut - m_Z^{4})
+ \frac{4s}{m_Z^{2}}
\biggr]
&\mbox{for}\ X=Y \\
\displaystyle
|tB_{t}^{ZZ}|^{2} \biggl[ \frac{4s}{t^{2}} + \frac{s}{m_Z^{4}} + \frac{4}{m_Z^{2}}\left(\frac{u}{t} - \frac{m_Z^{4}}{t^{2}}\right)
\biggr]
&\mbox{for}\ X\not=Y \\
\end{array}
\right..
\label{Eq:MtXY-ZZ-final}
\end{align}

Next, the u-channel contribution $|{\cal M}_{uXY}^{ZZ}|^{2}$ is given by
\begin{align}
|{\cal M}_{u}^{ZZ}|^{2}
&= \sum_{\text{spins}}
\epsilon^{*}_{\mu}(k_{1}) \epsilon^{*}_{\nu}(k_{2})
\epsilon_{\mu'}(k_{1}) \epsilon_{\nu'}(k_{2})
K_{u}^{\mu\nu} \bar{K}_{u}^{\mu'\nu'}
\nonumber\\
&= 
\left( - g_{\mu\mu'} + \frac{k_{1\mu} k_{1\mu'}}{k_{1}^{2}}\right)
\left( - g_{\nu\nu'} + \frac{k_{2\nu} k_{2\nu'}}{k_{2}^{2}}\right)
\sum_{\text{spins}}
K_{u}^{\mu\nu} \bar{K}_{u}^{\mu'\nu'}.
\label{Eq:MuXY-ZZ}
\end{align}
Substituting Eq.~(\ref{Eq:KuXY}) into Eq.~(\ref{Eq:MuXY-ZZ}),
we find
\begin{align}
|{\cal M}_{uXY}^{ZZ}|^{2}
&=
\left\{
\begin{array}{ll}
\displaystyle
|uA_{uX}^{ZZ}|^{2}
\biggl[
4 \left(\frac{t}{u} - \frac{m_Z^{4}}{u^{2}}\right) 
+ \frac{1}{m_Z^{4}} (ut - m_Z^{4})
+ \frac{4s}{m_Z^{2}}
\biggr]
&\mbox{for}\ X=Y \\
\displaystyle
|uB_{u}^{ZZ}|^{2} 
\biggl[
\frac{4s}{u^{2}} + \frac{s}{m_Z^{4}} + \frac{4}{m_Z^{2}}
\left(\frac{t}{u} - \frac{m_Z^{4}}{u^{2}} \right)
\biggr]
&\mbox{for}\ X\not=Y \\
\end{array}
\right..
\label{Eq:MuXY-ZZ-final}
\end{align}

Finally, the interference terms are given by
\begin{align}
&{\cal M}_{t}^{ZZ}{\cal M}_{u}^{ZZ\dag}
+ {\cal M}_{t}^{ZZ\dag} {\cal M}_{u}^{ZZ}
\nonumber\\
&= \sum_{\text{spins}}
\epsilon^{*}_{\mu}(k_{1}) \epsilon^{*}_{\nu}(k_{2})
\epsilon_{\mu'}(k_{1}) \epsilon_{\nu'}(k_{2})
(K_{t}^{\mu\nu} \bar{K}_{u}^{\mu'\nu'}
+ \bar{K}_{t}^{\mu\nu} K_{u}^{\mu'\nu'})
\nonumber\\
&= 
\left( - g_{\mu\mu'} + \frac{k_{1\mu} k_{1\mu'}}{k_{1}^{2}}\right)
\left( - g_{\nu\nu'} + \frac{k_{2\nu} k_{2\nu'}}{k_{2}^{2}}\right)
\sum_{\text{spins}}
(K_{t}^{\mu\nu} \bar{K}_{u}^{\mu'\nu'}
+ \bar{K}_{t}^{\mu\nu} K_{u}^{\mu'\nu'})
\label{Eq:MiXY-ZZ}
\end{align}
Substituting Eqs.~(\ref{Eq:KtXY}) and (\ref{Eq:KuXY})
into Eq.~(\ref{Eq:MiXY-ZZ}),
we find
\begin{align}
&{\cal M}_{tXY}^{ZZ}{\cal M}_{uXY}^{ZZ\dag}
+ {\cal M}_{tXY}^{ZZ\dag} {\cal M}_{uXY}^{ZZ}
\nonumber\\
&=
\left\{
\begin{array}{ll}
\displaystyle
ut (A_{tX}^{ZZ} A_{uX}^{ZZ*} + A_{uX}^{ZZ} A_{tX}^{ZZ*})
\cdot \biggl\{
 \frac{8m_Z^{2}s}{ut}
+ \frac{m_Z^{4} - ut}{m_Z^{4}}
- \frac{4s}{m_Z^{2}}
\biggr\} 
&\mbox{for}\ X=Y \\
\displaystyle
ut (B_{t}^{ZZ} B_{u}^{ZZ*} + B_{u}^{ZZ} B_{t}^{ZZ*}) \cdot \biggl\{
\frac{8s}{ut} + \frac{2s}{m_Z^{4}} - \frac{4(m_Z^{2}-u)(m_Z^{2}-t)}{ut}
\biggr\}
&\mbox{for}\ X\not=Y \\
\end{array}
\right..
\label{Eq:MiXY-ZZ-final}
\end{align}

From Eqs.~(\ref{Eq:MtXY-ZZ-final}), (\ref{Eq:MuXY-ZZ-final}), and
(\ref{Eq:MiXY-ZZ-final}), the total amplitudes of
$e_X^-e_{\bar{Y}}^+\to ZZ$ are given by for $X=Y$
\begin{align}
|{\cal M}^{ZZ}_{XY}|^{2}
&= |t A_{tX}^{ZZ}|^{2}\biggl\{
4 \left(\frac{u}{t} - \frac{m_Z^{4}}{t^{2}} \right)
+ \frac{1}{m_Z^{4}} (ut - m_Z^{4})
+ \frac{4s}{m_Z^{2}}
\biggr\}
\nonumber\\ & \quad
+ |u A_{uX}^{ZZ}|^{2} \biggl\{
4 \left(\frac{t}{u} - \frac{m_Z^{4}}{u^{2}} \right)
+ \frac{1}{m_Z^{4}} (ut - m_Z^{4})
+ \frac{4s}{m_Z^{2}}
\biggr\}
\nonumber\\ & \quad
+  2 \mbox{Re}[t A_{tX}^{ZZ}\cdot u A_{uX}^{ZZ*}] 
\biggl\{  
\frac{8 m_Z^{2}s}{ut} + \frac{m_Z^{4}-ut}{m_Z^{4}} 
- \frac{4s}{m_Z^{2}}
\biggr\};
\label{Eq:ee-to-ZZ-X=Y}
\end{align}
and for $X\not=Y$
\begin{align}
2|{\cal M}^{ZZ}_{XY}|^{2}
&= |tB_{t}^{ZZ}|^{2} \biggl\{
\frac{8s}{t^{2}} + \frac{2s}{m_Z^{4}} + \frac{8}{m_Z^{2}} 
\left(
\frac{u}{t} - \frac{m_Z^{4}}{t^{2}}
\right)
\biggr\}
\nonumber\\ & \quad
+ |uB_{u}^{ZZ}|^{2} \biggl\{
\frac{8s}{u^{2}} + \frac{2s}{m_Z^{4}} + \frac{8}{m_Z^{2}}
\left( \frac{t}{u} - \frac{m_Z^{4}}{u^{2}} \right)
\biggr\}
\nonumber\\& \quad
+ (2 \mbox{Re}[tB_{t}^{ZZ} \, uB_{u}^{ZZ*}]) \cdot
\biggl\{
\frac{8s}{ut}
+ \frac{2s}{m_Z^{4}}
- \frac{4(m_Z^{2} - t) (m_Z^{2}-u)}{ut}
\biggr\}.
\label{Eq:ee-to-ZZ-Xnot=Y}
\end{align}

\subsubsection{Cross section}

For the $e^-_Xe^+_{\bar{Y}}\to ZZ$ process, the cross section of the
initial states of the polarized electron and positron is given by
\begin{align}
\frac{d\sigma^{ZZ}}{d\cos\theta}(P_{e^-},P_{e^+},\cos\theta)
&=
\frac{1}{4}
\bigg\{
(1-P_{e^-})(1+P_{e^+})
\frac{d\sigma_{LL}^{ZZ}}{d\cos\theta}
+(1+P_{e^-})(1-P_{e^+})
\frac{d\sigma_{RR}^{ZZ}}{d\cos\theta}
\nonumber\\
&\ \ \ \
+(1-P_{e^-})(1-P_{e^+})
\frac{d\sigma_{LR}^{ZZ}}{d\cos\theta}
+(1+P_{e^-})(1+P_{e^+})
\frac{d\sigma_{RL}^{ZZ}}{d\cos\theta}
\bigg\},
\label{Eq:dsigma-ee-to-ZZ}
\end{align}
where $P_{e^-}$ and $P_{e^+}$ are the initial polarizations of the
electron and positron 
\begin{align}
\frac{d\sigma_{XY}^{ZZ}}{d\cos\theta}(\cos\theta):=
\frac{d\sigma}{d\cos\theta}(e_{X}^-e_{\bar{Y}}^+\to ZZ)
=\frac{\beta_Z}{32\pi s} {|{\cal M}^{ZZ}_{XY}|^{2}},
\label{Eq:dsigma-ee-to-ZZ-XY}
\end{align}
where 
${\cal M}_{XY}^{ZZ}$ are given in Eqs.~(\ref{Eq:ee-to-ZZ-X=Y}) and 
(\ref{Eq:ee-to-ZZ-Xnot=Y})

The total cross section of the $e^-e^+\to ZZ$ process with the initial
polarizations can be defined by integrating the
differential cross section in Eq.~(\ref{Eq:dsigma-ee-to-ZZ}) with the
angle $\theta$
\begin{align}
 \sigma^{ZZ}(P_{e^-},P_{e^+}):=\frac{1}{2}
 \int_{-1}^{1}
 \frac{d\sigma^{ZZ}}{d\cos\theta}(P_{e^-},P_{e^+},\cos\theta)
 d\cos\theta,
\end{align}
where the minimum and maximal values of $\cos\theta$ are determined
by each detector and we cannot use date near $\cos\theta\simeq \pm1$.
The total cross section of $e^-e^+\to ZZ$ is given by
\begin{align}
\sigma^{ZZ}(P_{e^-},P_{e^+})
&=\frac{1}{4}
(1-P_{e^-})(1+P_{e^+})
\sigma_{LL}^{ZZ}
+\frac{1}{4}
(1+P_{e^-})(1-P_{e^+})
\sigma_{RR}^{ZZ}
\nonumber\\
&\ \ \
+\frac{1}{4}
(1-P_{e^-})(1-P_{e^+})
\sigma_{LR}^{ZZ}
+\frac{1}{4}
(1+P_{e^-})(1+P_{e^+})
\sigma_{RL}^{ZZ},
\label{Eq:sigma-total-ZZ}
\end{align}
where 
\begin{align}
\sigma_{XY}^{ZZ}&:=\frac{1}{2}
 \int_{-1}^{1}
 \frac{d\sigma_{XY}^{ZZ}}{d\cos\theta}(\cos\theta)
 d\cos\theta,
\label{Eq:sigma-total-ZZ}
\end{align}
where $X=L,R$; $d\sigma_{XY}^{ZZ}/d\cos\theta$ 
are given in Eq.~(\ref{Eq:dsigma-ee-to-ZZ-XY}).

The statistical error of the cross section 
$\sigma^{ZZ}(P_{e^-},P_{e^+})$ is given by
\begin{align}
\Delta \sigma^{ZZ}(P_{e^-},P_{e^+})=
 \frac{\sigma^{ZZ}(P_{e^-},P_{e^+})}
 {\sqrt{N^{ZZ}}},\ \ 
 N^{ZZ} =  L_{\rm int} \cdot
 \sigma^{ZZ}(P_{e^-},P_{e^+}),
\label{Eq:stat-error-sigma-ZZ}
\end{align}
where $L_{\rm int}$ is an integrated luminosity, and 
$N^{ZZ}$ is the number of events for the $e^-e^+\to ZZ$ process.
Note that the $Z$ boson cannot be observed directly, so
we need to choose the decay modes of the $Z$ boson, and then the
available number of events must be $N^{ZZ}$ multiplied by the branching
ratio of each selected decay mode. The amount of the deviation of the
cross section of the $e^-e^+\to ZZ$ process from the SM in the GHU model
is given by
\begin{align}
\Delta_\sigma^{ZZ}(P_{e^-},P_{e^+}):=
\frac{
\left[\sigma^{ZZ}(P_{e^-},P_{e^+})\right]_{\rm GHU}}
{\left[\sigma^{ZZ}(P_{e^-},P_{e^+})\right]_{\rm SM}}-1,
\label{Eq:Delta_sigma-ZZ}
\end{align}
where $\left[\sigma^{ZZ}(P_{e^-},P_{e^+})\right]_{\rm GHU}$ 
and $\left[\sigma^{ZZ}(P_{e^-},P_{e^+})\right]_{\rm SM}$ 
stand for the cross sections of the $e^-e^+\to ZZ$ process in
the SM and the GHU model, respectively.
The same notation is used for other cases in the followings.

\subsubsection{Left-right asymmetry}

We define an observable  left-right asymmetry of the $e^-e^+\to ZZ$
process as 
\begin{align}
A_{LR}^{ZZ}(P_{e^-},P_{e^+})
:=\frac{\sigma^{ZZ}(P_{e^-},P_{e^+})-\sigma^{ZZ}(-P_{e^-},-P_{e^+})}
{\sigma^{ZZ}(P_{e^-},P_{e^+})+\sigma^{ZZ}(-P_{e^-},-P_{e^+})}
\label{Eq:ALR-ZZ-def}
\end{align}
for $P_{e^-}<0$ and $|P_{e^-}|>|P_{e^+}|$.

The statistical error of the left-right asymmetry is given by
\begin{align}
 \Delta A_{LR}^{ZZ}(P_{e^-},P_{e^+})&=
 2\frac{\sqrt{N_{L}^{ZZ}N_{R}^{ZZ}}
 \left(\sqrt{N_{L}^{ZZ}}+\sqrt{N_{R}^{ZZ}}\right)}
 {(N_{L}^{ZZ}+N_{R}^{ZZ})^2},
\label{Eq:Error-A_LR-ZZ}
\end{align}
where $N_{L}^{ZZ}=L_{\rm int} \, \sigma^{ZZ}(P_{e^-},P_{e^+})$
and $N_{R}^{ZZ}=L_{\rm int} \, \sigma^{ZZ}(-P_{e^-},-P_{e^+})$
are the numbers of the events 
for $P_{e^-}<0$ and $|P_{e^-}|>|P_{e^+}|$.
The amount of the deviation from the SM in Eq.~(\ref{Eq:ALR-ZZ-def})
is characterized by
\begin{align}
\Delta_{A_{LR}}^{ZZ}(P_{e^-},P_{e^+})&:=
 \frac{\left[A_{LR}^{ZZ}(P_{e^-},P_{e^+})\right]_{\rm GHU}}
{\left[A_{LR}^{ZZ}(P_{e^-},P_{e^+})\right]_{\rm SM}}-1.
\label{Eq:Delta_A_LR-ZZ}
\end{align}

\section{Numerical analysis}
\label{Sec:Results}

We analyze cross sections of $W$ and $Z$ boson pair production
processes $e^-e^+\to W^-W^+$ and $e^-e^+\to ZZ$
for the initial states of unpolarized and polarized electrons and
positrons,
where we use the formula of the cross sections given in
Sec.~\ref{Sec:Cross-section}. For the values of the initial
polarizations, we mainly use $(P_{e^-},P_{e^+})=(\mp0.8,\pm0.3)$
with the ILC experiment in mind. 

\subsection{$e^-e^+\to W^-W^+$}
\label{Sec:ee-to-WW}

Here we evaluate observables of the $e^-e^+\to W^-W^+$ process in the SM
and the GHU model at tree level. We use the
parameter sets A$_{\pm}$, B$_{\pm}$, C$_{\pm}$ listed in
Tables~\ref{Table:Parameter-sets}, 
\ref{Table:Gauge-Charged-Lepton-Couplings},
\ref{Table:Triple-Gauge-Couplings}.

\begin{figure}[htb]
\begin{center}
\includegraphics[bb=0 0 363 240,height=5cm]{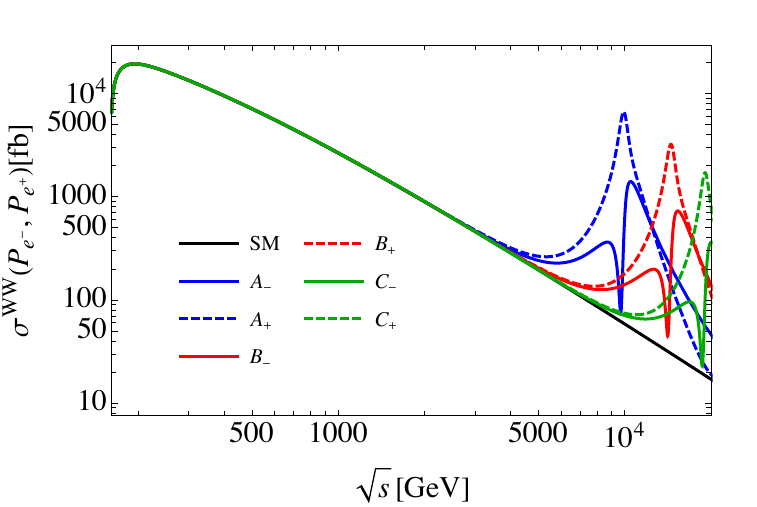}
\includegraphics[bb=0 0 363 240,height=5cm]{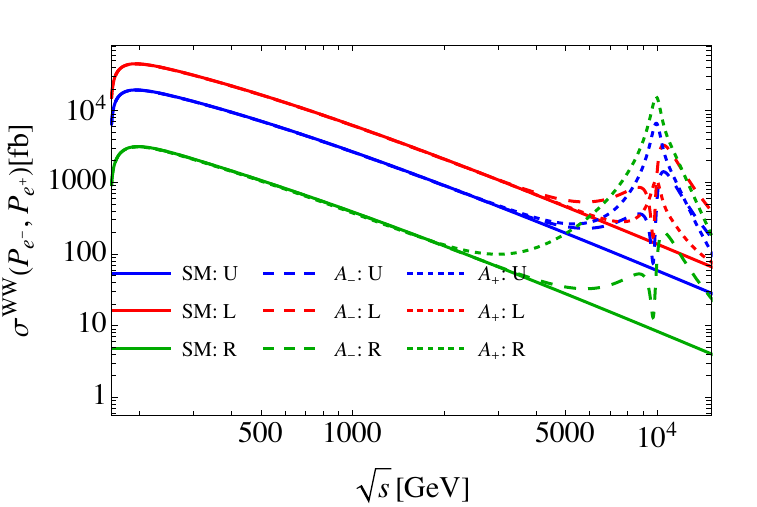}
 \caption{\small
 The total cross sections of the $e^-e^+\to W^-W^+$ process
 in the SM and the GHU model are shown in wider range of $\sqrt{s}$.
 The left figure shows
 the $\sqrt{s}$ dependence of 
 $\sigma^{WW}(P_{e^-}=0,P_{e^+}=0)$ in the SM and the GHU model
 with unpolarized electron and positron beams, where $A_\pm$, $B_\pm$,
 $C_\pm$ are 
 the names of the parameter sets listed in  
 Table~\ref{Table:Parameter-sets}.
 The right figure shows 
 the $\sqrt{s}$ dependence of 
 $\sigma^{WW}(P_{e^-},P_{e^+})$ in the SM and the GHU model whose
 parameter set is $A_-$ 
 with the three different polarizations U, L, R, where 
 U, L, R stand for $(P_{e^-},P_{e^+})=(0,0),(-0.8,+0.3),(+0.8,-0.3)$,
 respectively.
 }
 \label{Figure:sigma-WW}
\end{center}
\end{figure}

In Figure~\ref{Figure:sigma-WW}, we show the total cross sections of the
$e^-e^+\to W^-W^+$ process  in the SM and the GHU model in wider range
of $\sqrt{s}$ with unpolarized and polarized electron and positron
beams, where $A_\pm$, $B_\pm$, $C_\pm$ are  the names of the parameter
sets listed in Table~\ref{Table:Parameter-sets}.
From Figure~\ref{Figure:sigma-WW}, we can see that the deviation from
the SM of the GHU 
model is very large due to the effect of resonances around the mass
scale of the $Z'$ bosons.
We focus on the cross sections around $\sqrt{s}\simeq 10$\,TeV for
$A_\pm$.
Three $Z'$ bosons $\gamma^{(1)}$, $Z^{(1)}$, and $Z_R^{(1)}$ contribute
to this resonance via the s-channel process, where this system take
place very strong cancellation between s-channel, t-channel, and
interference terms. The mass of $Z_R^{(1)}$ is slightly smaller than the
masses of $\gamma^{(1)}$ and $Z^{(1)}$ and the sign of the coupling
constants to the electrons and $W$ bosons varies with the sign of the
bulk mass parameters given in
Tables~\ref{Table:Gauge-Charged-Lepton-Couplings} and
\ref{Table:Triple-Gauge-Couplings}, and Table 8 in
Ref.~\cite{Funatsu:2023jng},
For $A_pm$, the coupling constants of the $Z'$ bosons to 
electrons and $W$ bosons are given in
Tables~\ref{Table:KK-mode-couplings} and
\ref{Table:KK-mode-couplings-positive}.
In the $A_-$ case, the combination is caused by interference effects
among the SM gauge bosons and $Z'$ bosons at energies slightly below
the mass of $Z_R^{(1)}$. In the case of $A_+$, on the other hand, the
sign of the coupling constants is such that no interference occurs.

\begin{figure}[htb]
\begin{center}
\includegraphics[bb=0 0 363 230,height=5cm]{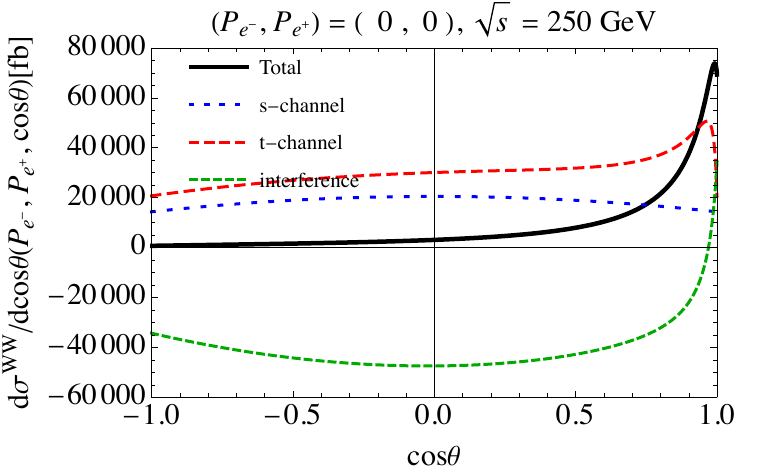}
\includegraphics[bb=0 0 363 230,height=5cm]{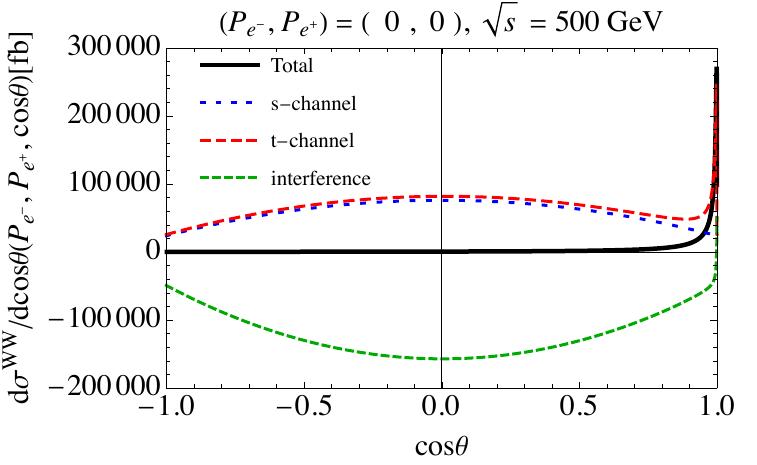}
 \caption{\small
 The differential cross sections of the $e^-e^+\to W^-W^+$ process
 in the SM 
 $\frac{d\sigma^{WW}}{d\cos\theta}(P_{e^-},P_{e^+},\cos\theta)$
 are shown at $\sqrt{s}=250$\,GeV and $500$\,GeV for the
 left and right figures, respectively.
 Total stands for differential cross section including all the
 contribution from s-channel, t-channel, and interference terms.
 s-channel, t-channel, and interference stand for 
 for differential cross section only including each contribution.
 }
 \label{Figure:dsigma-WW}
\end{center}
\end{figure}

In Figure~\ref{Figure:dsigma-WW}, we show the angular distribution 
of the cross sections of the $e^-e^+\to W^-W^+$ process 
in the SM with $(P_{e^-},P_{e^+})=(0,0)$ at $\sqrt{s}=250$\,GeV and
$500$\,GeV with unpolarized $e^\pm$ beams,
where $\sigma^{WW}(P_{e^-},P_{e^+})$ given in
Eq.~(\ref{Eq:dsigma-ee-to-WW}).
Figure~\ref{Figure:dsigma-WW} shows that except for
$\cos\theta\simeq 1$, there is very strong cancellation among the
s-channel, t-channel, and interference terms. 
Furthermore, the larger $\sqrt{s}$, the more the forward cross section
increases while the backward cross section cancels out more strongly.

\begin{figure}[htb]
\begin{center}
\begin{tabular}{|c|c|c|}
\hline
\rowcolor[gray]{0.9}
$(P_{e^-},P_{e^-})$
&$(\sqrt{s},L_{\rm int})=(250\,\mbox{GeV},1\,\mbox{ab}^{-1})$
&$(\sqrt{s},L_{\rm int})=(500\,\mbox{GeV},2\,\mbox{ab}^{-1})$\\
\hline
\raisebox{2cm}{$(-0.8,+0.3)$}
&\includegraphics[bb=0 0 363 230,height=4cm]{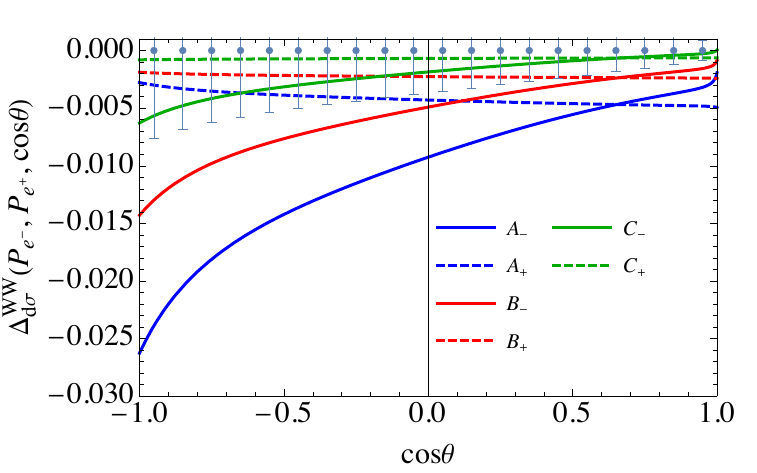}
&\includegraphics[bb=0 0 363 230,height=4cm]{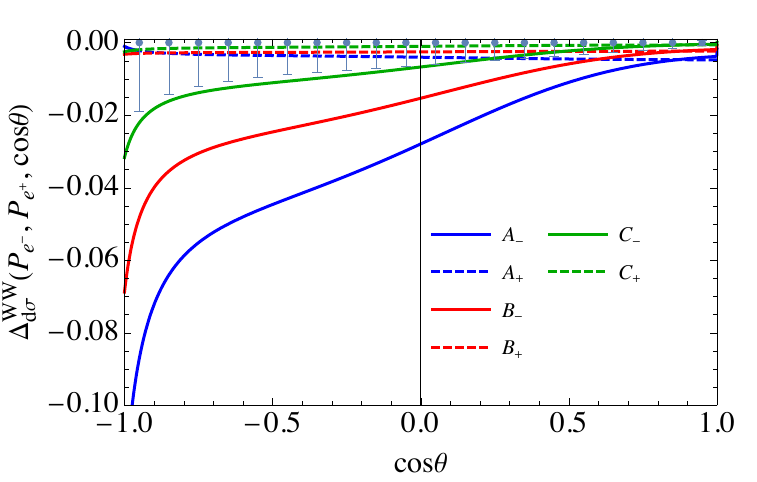}\\
\hline
\raisebox{2cm}{$(0,0)$}
&\includegraphics[bb=0 0 363 230,height=4cm]{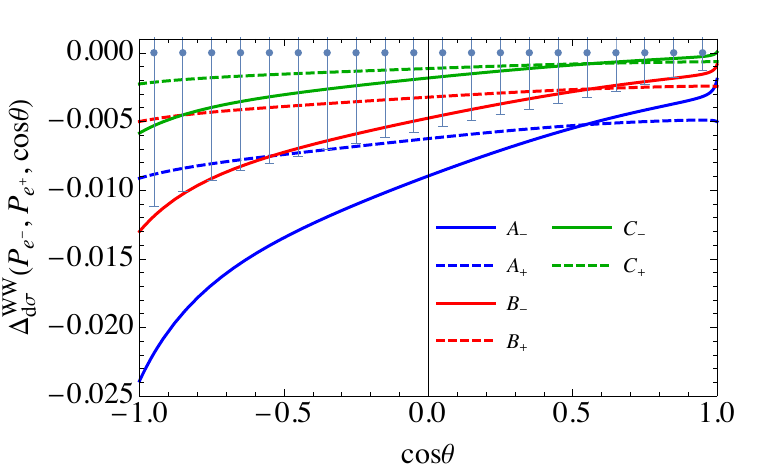}
&\includegraphics[bb=0 0 363 230,height=4cm]{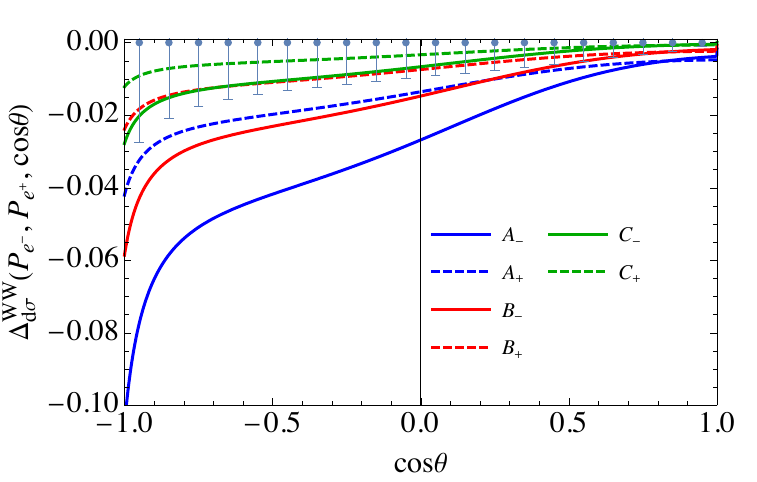}\\
\hline
\raisebox{2cm}{$(+0.8,-0.3)$}
&\includegraphics[bb=0 0 363 230,height=4cm]{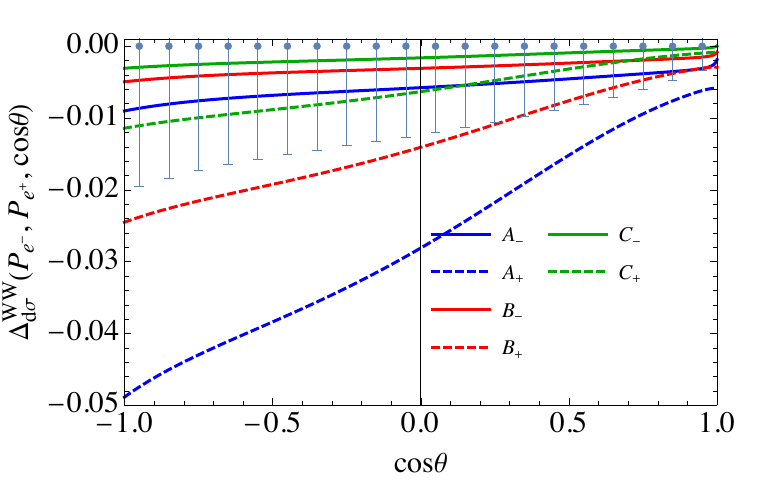}
&\includegraphics[bb=0 0 363 230,height=4cm]{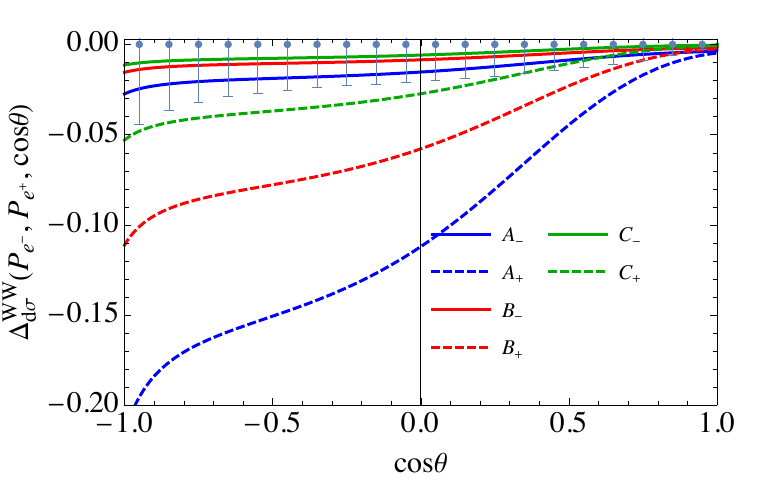}\\
\hline
\end{tabular}
 \caption{\small
 The deviations from the SM in the GHU for the $e^-e^+\to W^-W^+$
 process 
 $\Delta_{d\sigma}^{WW}(P_{e^-},P_{e^+},\cos\theta)$
 are shown with
 $(P_{e^-},P_{e^+})=(-0.8,+0.3),(0,0),(+0.8,-0.3)$
 for upper, middle and lower rows, respectively,
 where 
 $\Delta_{d\sigma}^{WW}(P_{e^-},P_{e^+},\cos\theta):=
 \left[\frac{d\sigma^{WW}}{d\cos\theta}
 (P_{e^-},P_{e^+},\cos\theta)\right]_{\rm GHU}
 \left[\frac{d\sigma^{WW}}{d\cos\theta}
 (P_{e^-},P_{e^+},\cos\theta)\right]_{\rm SM}^{-1}-1$.
 The left and right side figures show the deviation at
 $\sqrt{s}=250$\,GeV and $500$\,GeV, respectively. 
 The error bars represent statistical errors in the SM 
 at $\sqrt{s}=250 \,$GeV with 1\,ab$^{-1}$ data 
 and at $\sqrt{s}=500\,$GeV with 2$\,$ab$^{-1}$ data 
 by using leptonic decays $W^{\pm}\to \ell^\pm\nu$.
 The branching ratio of 
 $W^+W^-\to \ell^-\ell^{\prime+}\nu\nu'$ is 10.615\%
 since $\mbox{Br}(W^+\to\ell^+\nu)=(10.86\pm 0.09)$\%
 \cite{ParticleDataGroup:2022pth}.
 Each bin is given by $\cos\theta=[k-0.05,k+0.05]$ 
 ($k=-0.95,-0.85,\cdots,0.95$).
 }
 \label{Figure:Delta-dsigma-WW}
\end{center}
\end{figure}

In Figure~\ref{Figure:Delta-dsigma-WW}, the deviations from the SM  in
the GHU with polarized $e^\pm$ beams are shown.
The deviation from the SM in the GHU
model with unpolarized and left- and right-handed polarized $e^\pm$ beams
$(P_{e^-},P_{e^+})=(0,0),(-0.8,+0.3),(+0.8,-0.3)$, respectively.
The statistical errors in the SM are estimated
by using leptonic decays $W^{\pm}\to \ell^\pm\nu$.
Since $\mbox{Br}(W^+\to\ell^+\nu)=(10.86\pm 0.09)$\%
\cite{ParticleDataGroup:2022pth},
the branching ratio of 
$W^+W^-\to \ell^-\ell^{\prime+}\nu\nu'$ is 10.615\%.
The estimates from the statistical errors in this figure show that the
exploration area for the GHU model is wider for $\sqrt{s}=500$\,GeV,
$L_{\rm int}=1$ab$^{-1}$ than for 
$\sqrt{s}=250$\,GeV, $L_{\rm int}=2$ab$^{-1}$.
We also find that for the $A_-$, $B_-$, $C_-$ cases, where the bulk
masses are negative, the deviation from the SM is larger in the
left-handed $e^\pm$ beams because the left-handed coupling constants are
larger than the right-handed ones, while for the $A_+$, $B_+$, $C_+$
cases, where the bulk masses are positive, the deviation from the SM is
larger in the right-handed  $e^\pm$ beams because the right-handed
coupling constants are larger than the left-handed ones.

\begin{figure}[htb]
\begin{center}
\includegraphics[bb=0 0 363 232,height=5cm]{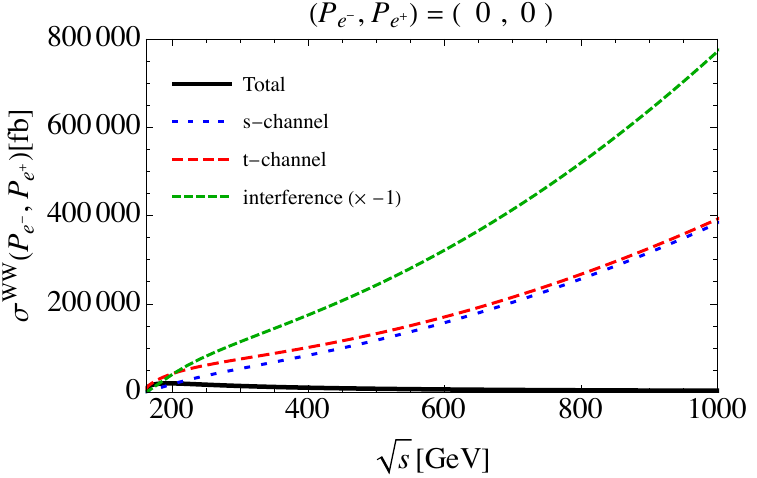}
\ \ 
\includegraphics[bb=0 0 363 236,height=5cm]{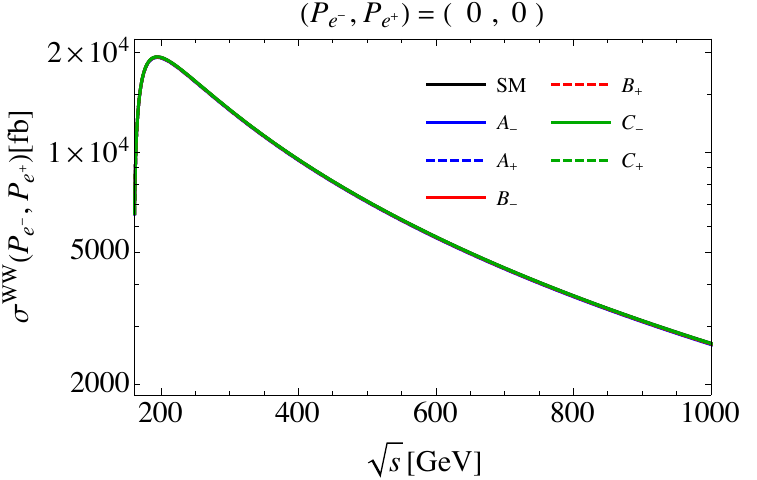}
 \caption{\small
 The total cross sections of the $e^-e^+\to W^-W^+$ process
 in the SM and the GHU model
 $\sigma^{WW}(P_{e^-},P_{e^+})$
 are shown in range of
 $\sqrt{s}=[165,1000]$\,GeV.
 The left figure shows
 the $\sqrt{s}$ dependence of 
 $\sigma^{WW}(P_{e^-}=0,P_{e^+}=0)$ in the SM 
 with unpolarized electron and positron beams, 
 Total stands for differential cross section including all the
 contribution from s-channel, t-channel, and interference terms.
 s-channel, t-channel, and interference stand for
 for cross section only including each contribution.
 The right figure shows the $\sqrt{s}$ dependence of 
 $\sigma^{WW}(P_{e^-},P_{e^+})$ in the SM and the GHU model whose
 parameters set are $A_\pm$, $B_\pm$, $C_\pm$.
 }
 \label{Figure:sigma-WW-low}
\end{center}
\end{figure}

From Figure~\ref{Figure:sigma-WW-low}, 
the total cross sections of the $e^-e^+\to W^-W^+$ process
in the SM and the GHU model are shown in range of
$\sqrt{s}=[165,1000]$\,GeV.
The left figure shows
the $\sqrt{s}$ dependence of 
$\sigma^{WW}(P_{e^-}=0,P_{e^+}=0)$ in the SM 
with unpolarized electron and positron beams, 
Total stands for differential cross section including all the
contribution from s-channel, t-channel, and interference terms.
s-channel, t-channel, and interference stand for
for cross section only including each contribution.
The right figure shows the $\sqrt{s}$ dependence of 
$\sigma^{WW}(P_{e^-},P_{e^+})$ in the SM and the GHU model whose
parameters set are $A_\pm$, $B_\pm$, $C_\pm$.
From the left figure, we can see that for larger $\sqrt{s}$, stronger
cancellation between s-channel, t-channel, and interference terms is
taking place. This figure shows the SM case, but the same phenomenon
occurs in the GHU model.
From the right figure, we find that the cross sections in the GHU with 
the parameter sets $A_\pm$, $B_\pm$, $C_\pm$ are almost the same as that
in the SM.

\begin{figure}[htb]
\begin{center}
\includegraphics[bb=0 0 363 241,height=3.4cm]{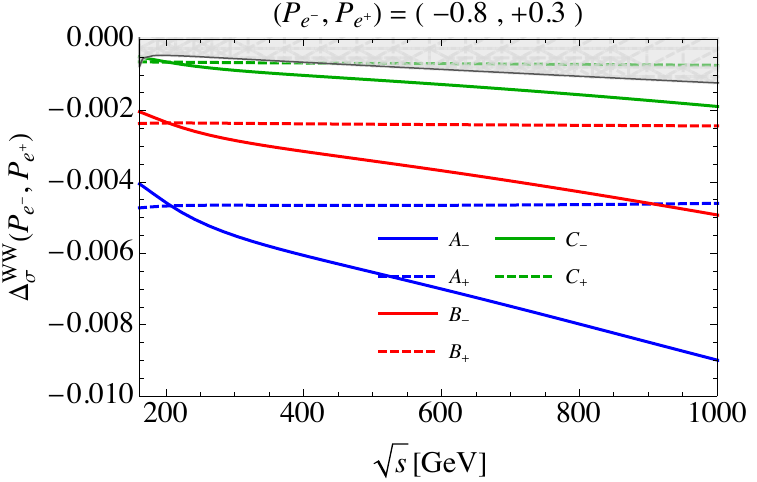}
\includegraphics[bb=0 0 363 241,height=3.4cm]{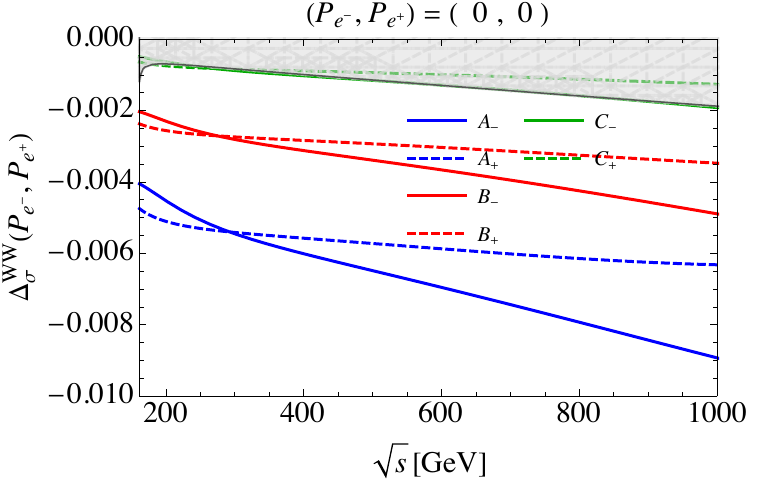}
\includegraphics[bb=0 0 363 241,height=3.4cm]{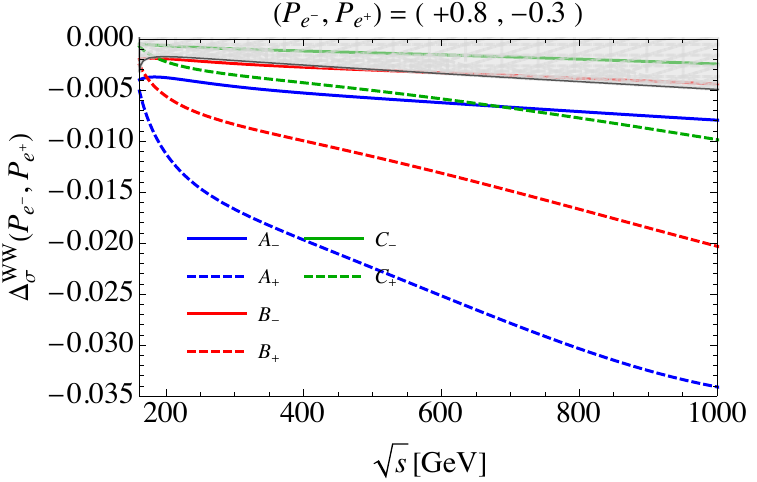}
 \caption{\small
 The $\sqrt{s}$ dependence of the deviation from the SM in
 the GHU models for the total cross sections,
 $\Delta_{\sigma}^{WW}(P_{e^-},P_{e^+})$,
 is shown for $(P_{e^-},P_{e^+})=(-0.8,+0.3),(0,0),(+0.8,-0.3)$,
 respectively,
 where 
 $\Delta_\sigma^{WW}(P_{e^-},P_{e^+}):=
 \left[\sigma^{WW}(P_{e^-},P_{e^+})\right]_{\rm GHU}/
 \left[\sigma^{WW}(P_{e^-},P_{e^+})\right]_{\rm SM}-1$.
 The gray region represents the $1\sigma$ statistical error
 estimated  by using leptonic decays $W^{\pm}\to \ell^\pm\nu$
 and the integrated luminosity  $L_{\rm int}=1\,\mbox{ab}^{-1}$.
 } 
 \label{Figure:Delta-sigma-WW}
\end{center}
\end{figure}

In Figure~\ref{Figure:Delta-sigma-WW}, the $\sqrt{s}$
dependence of the deviation from the SM in  the GHU models for the total
cross sections, $\Delta_{\sigma}^{WW}(P_{e^-},P_{e^+})$, is shown, where
$\Delta_\sigma^{WW}(P_{e^-},P_{e^+})$ is given in
Eq.~(\ref{Eq:Delta_sigma-WW}).
The $1\sigma$ statistical errors are estimated 
by using the total cross section of $W$ boson pair production.
We find that for the $A_-$, $B_-$, $C_-$ cases, where the bulk masses are
negative, the deviation from the SM is larger in the 
left-handed $e^\pm$ beams, while for the $A_+$, $B_+$, $C_+$
cases, where the bulk masses are positive, the deviation from the SM is
larger in the right-handed  $e^\pm$ beams.

\begin{figure}[htb]
\begin{center}
\includegraphics[bb=0 0 363 240,height=5cm]{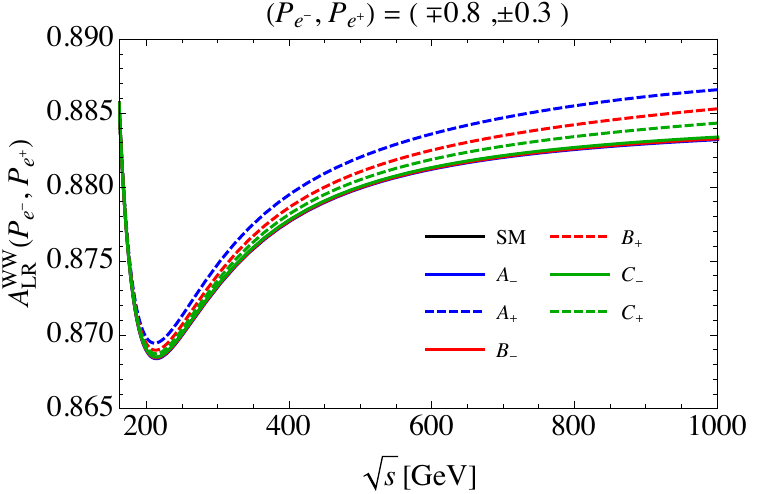}
\includegraphics[bb=0 0 363 235,height=5cm]{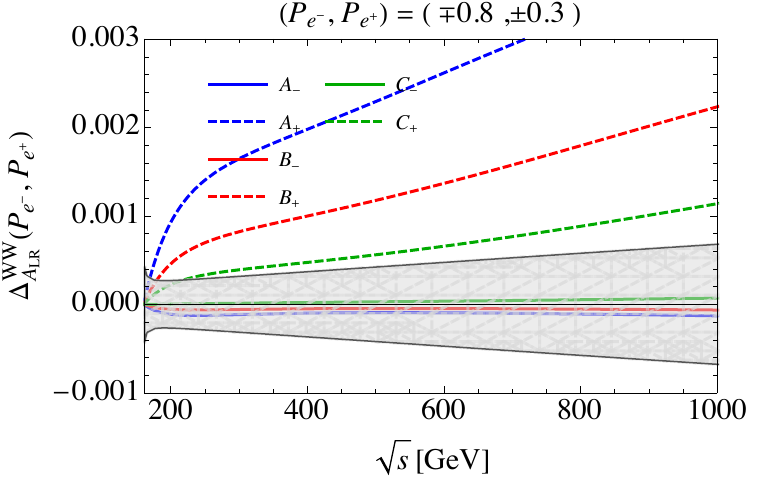}
 \caption{\small
 The $\sqrt{s}$ dependence of the left-right asymmetry of 
 the $e^-e^+\to W^-W^+$ process and the deviation from the SM  are shown. 
 The left figure shows the $\sqrt{s}$ dependence of
 $A_{LR}^{WW}(P_{e^-},P_{e^+})$ in the SM and the GHU model.
 The right figure shows the $\sqrt{s}$ dependence of
 $\Delta_{A_{LR}}^{WW}(P_{e^-},P_{e^+})$, 
 where  
 $\Delta_{A_{LR}}^{WW}(P_{e^-},P_{e^+}):=
 \left[A_{LR}^{WW}(P_{e^-},P_{e^+})\right]_{\rm GHU}/
 \left[A_{LR}^{WW}(P_{e^-},P_{e^+})\right]_{\rm SM}-1$.
 The energy ranges $\sqrt{s}$ in the first and second figures are
 $\sqrt{s}=[200,3000] \,$GeV, $\sqrt{s}=[200,1000] \,$GeV, respectively.
 The gray region represents the $1\sigma$ statistical error in the SM 
 at each $\sqrt{s}$ with $1$$\,$ab$^{-1}$ for 
 each polarized initial states
 $(P_{e^-},P_{e^+})=(\mp0.8,\pm0.3)$
 by using the total cross section of $W$ boson pair production.
 } 
 \label{Figure:ALR-WW}
\end{center}
\end{figure}

In Figure~\ref{Figure:ALR-WW},
the $\sqrt{s}$ dependence of the left-right asymmetry $A_{LR}^{WW}$
of the $e^-e^+\to W^-W^+$ processes and the deviation from the SM
$\Delta_{A_{LR}}^{WW}$ are shown,
where $A_{LR}^{WW}$ and $\Delta_{A_{LR}}^{WW}$ are 
given in Eqs.~(\ref{Eq:ALR-WW-def}) and (\ref{Eq:Delta_A_LR-WW}),
respectively.
The $1\sigma$ statistical error in the SM 
at each $\sqrt{s}$ with $1$$\,$ab$^{-1}$ for 
each polarized initial electron and positron
$(P_{e^-},P_{e^+})=(\mp0.8,\pm0.3)$
is estimated  by using the total cross section of $W$ boson pair
production and the branching ratio of $W^{\pm}\to \ell^\pm\nu$.
From Figure~\ref{Figure:ALR-WW}, we find that by using 
the left-right asymmetry $A_{LR}^{WW}$ 
we can explore higher KK scales than the constraints from the LHC
experiment only for the positive bulk mass.

\subsection{$e^-e^+\to ZZ$}
\label{Sec:ee-to-ZZ}

Here we evaluate observables of the $e^-e^+\to ZZ$ process in the SM
and the GHU model at tree level. As the same in Sec.~\ref{Sec:ee-to-WW},
we use the
parameter sets A$_{\pm}$, B$_{\pm}$, C$_{\pm}$ listed in
Tables~\ref{Table:Parameter-sets}, 
\ref{Table:Gauge-Charged-Lepton-Couplings},
\ref{Table:Triple-Gauge-Couplings}.

\begin{figure}[htb]
\begin{center}
\includegraphics[bb=0 0 363 240,height=5cm]{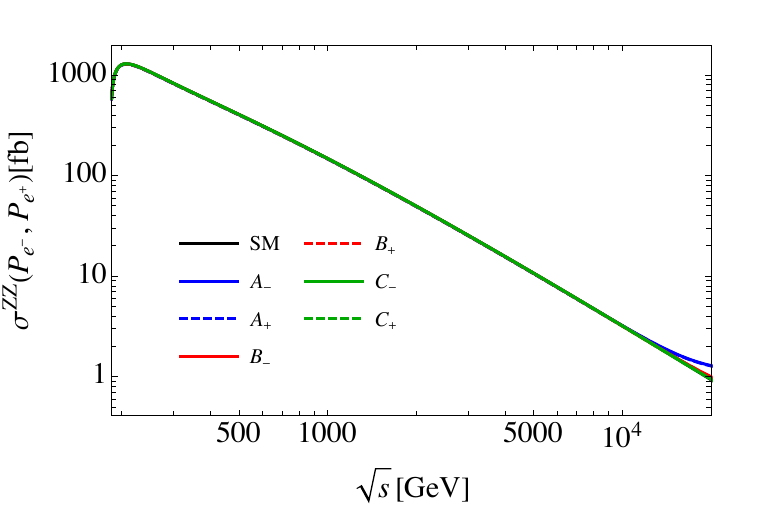}
\includegraphics[bb=0 0 363 240,height=5cm]{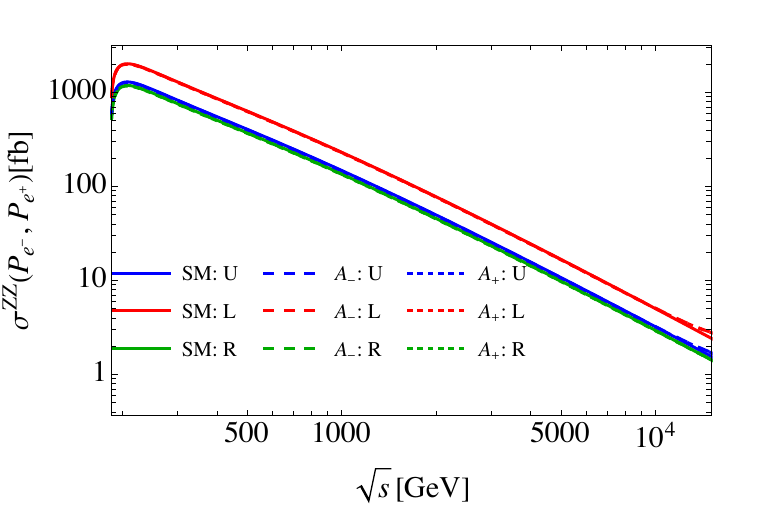}
 \caption{\small
 The total cross sections of the $e^-e^+\to ZZ$ process
 in the SM and the GHU model 
 $\sigma^{ZZ}(P_{e^-},P_{e^+})$
 are shown in wider range of $\sqrt{s}$.
 The left figure shows
 the $\sqrt{s}$ dependence of 
 $\sigma^{ZZ}(P_{e^-}=0,P_{e^+}=0)$ in the SM and the GHU model
 with unpolarized electron and positron beams, where $A_\pm$, $B_\pm$,
 $C_\pm$ are 
 the names of the parameter sets listed in  
 Table~\ref{Table:Parameter-sets}.
 The right figure shows 
 the $\sqrt{s}$ dependence of 
 $\sigma^{ZZ}(P_{e^-},P_{e^+})$ in the SM and the GHU model whose
 parameter sets are $A_\pm$ 
 with the three different polarizations U, L, R, where 
 U, L, R stand for $(P_{e^-},P_{e^+})=(0,0),(-0.8,+0.3),(+0.8,-0.3)$,
 respectively.
 }
 \label{Figure:sigma-ZZ}
\end{center}
\end{figure}

In Figure~\ref{Figure:sigma-ZZ}, we show the total cross sections of the
$e^-e^+\to ZZ$ process  in the SM and the GHU model in wider range
of $\sqrt{s}$ with unpolarized and polarized electron and positron
beams.
Unlike the $e^-e^+\to W^-W^+$ process, the $e^-e^+\to ZZ$ process has
only t- and u-channel contributions, so there are no resonances even
near the $Z'$ bosons.

\begin{figure}[htb]
\begin{center}
\includegraphics[bb=0 0 363 230,height=5cm]{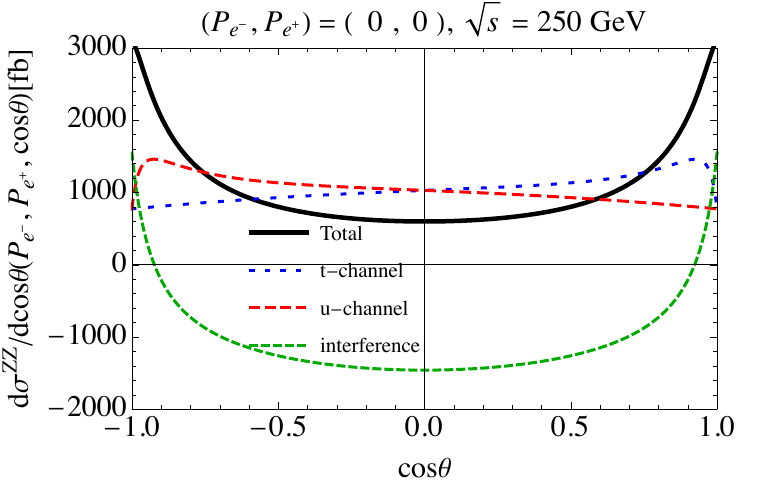}
\includegraphics[bb=0 0 363 230,height=5cm]{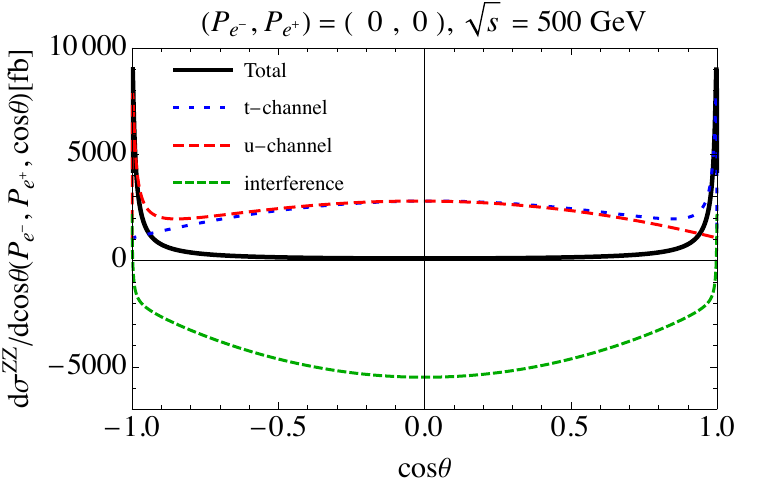}
 \caption{\small
 The differential cross sections of the $e^-e^+\to ZZ$ process
 in the SM  with unpolarized electron and positron beams
 $\frac{d\sigma^{ZZ}}{d\cos\theta}(P_{e^-},P_{e^+},\cos\theta)$
 are shown at $\sqrt{s}=250$\,GeV and  $500$\,GeV for
 the left and right figures, respectively.
 Total stands for differential cross section including all the
 contribution from s-channel, t-channel, and interference terms.
 s-channel, t-channel, and interference stand for
 for cross section only including each contribution.
 }
 \label{Figure:dsigma-ZZ}
\end{center}
\end{figure}

In Figure~\ref{Figure:dsigma-ZZ}, we show the angular distribution 
of the differential cross sections of the $e^-e^+\to W^-W^+$ process at 
$\sqrt{s}=250$\,GeV and $500$\,GeV with unpolarized $e^\pm$ beams
$(P_{e^-},P_{e^+})=(0,0)$.
Figure~\ref{Figure:dsigma-ZZ} shows that except for
$\cos\theta\simeq \pm1$, there is very strong cancellation among the
t-channel, u-channel, and interference terms. 
Furthermore, the larger $\sqrt{s}$, the more the forward and backward
cross sections increase.

\begin{figure}[htb]
\begin{center}
\begin{tabular}{|c|c|c|}
\hline
\rowcolor[gray]{0.9}
$(P_{e^-},P_{e^-})$
&$(\sqrt{s},L_{\rm int})=(250\,\mbox{GeV},1\,\mbox{ab}^{-1})$
&$(\sqrt{s},L_{\rm int})=(500\,\mbox{GeV},2\,\mbox{ab}^{-1})$\\
\hline
\raisebox{2cm}{$(-0.8,+0.3)$}
&\includegraphics[bb=0 0 363 230,height=4cm]{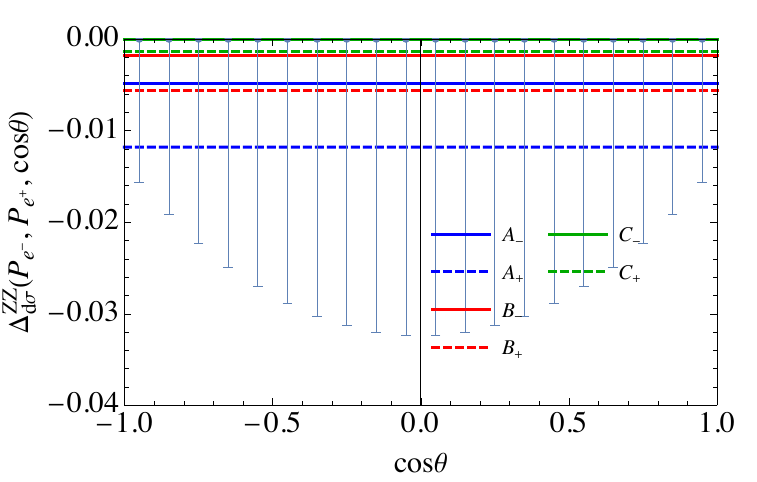}
&\includegraphics[bb=0 0 363 230,height=4cm]{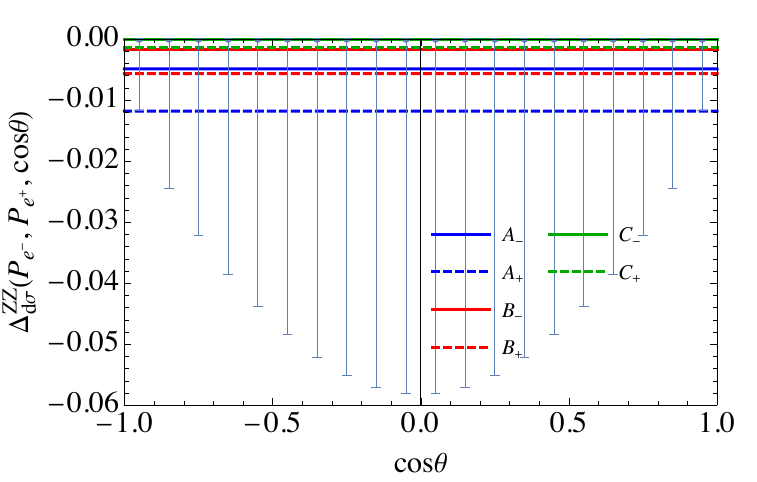}\\
\hline
\raisebox{2cm}{$(0,0)$}
&\includegraphics[bb=0 0 363 230,height=4cm]{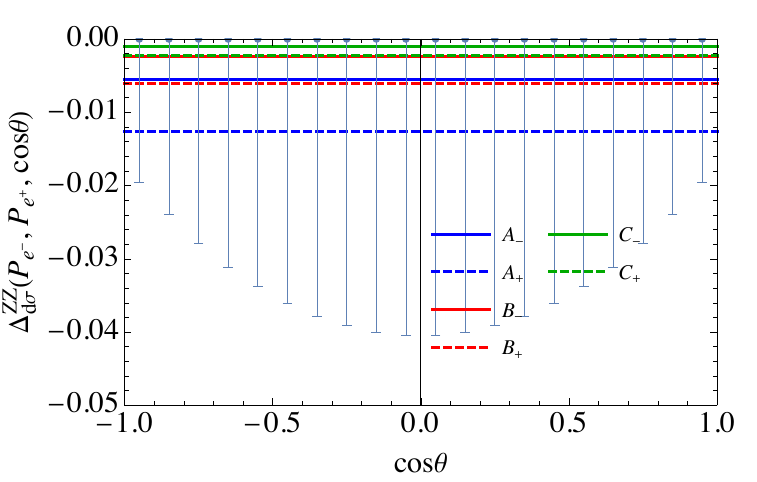}
&\includegraphics[bb=0 0 363 230,height=4cm]{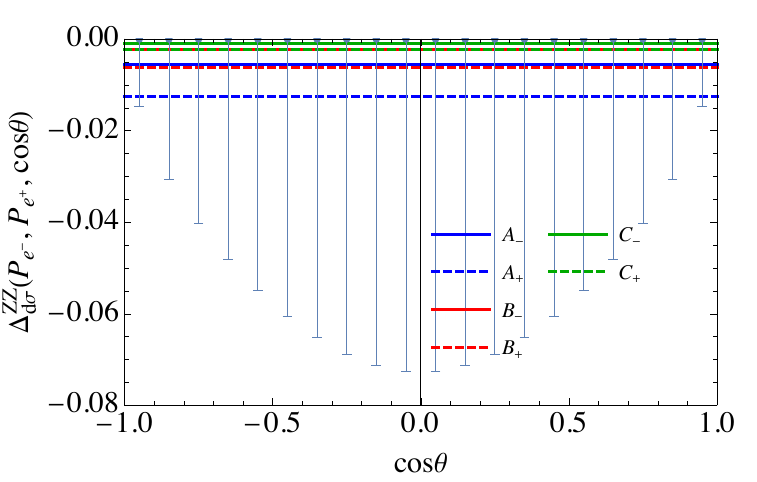}\\
\hline
\raisebox{2cm}{$(+0.8,-0.3)$}
&\includegraphics[bb=0 0 363 230,height=4cm]{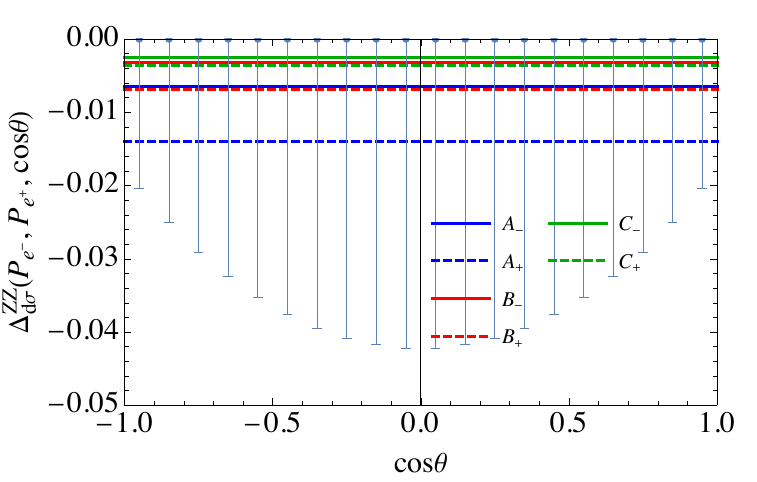}
&\includegraphics[bb=0 0 363 230,height=4cm]{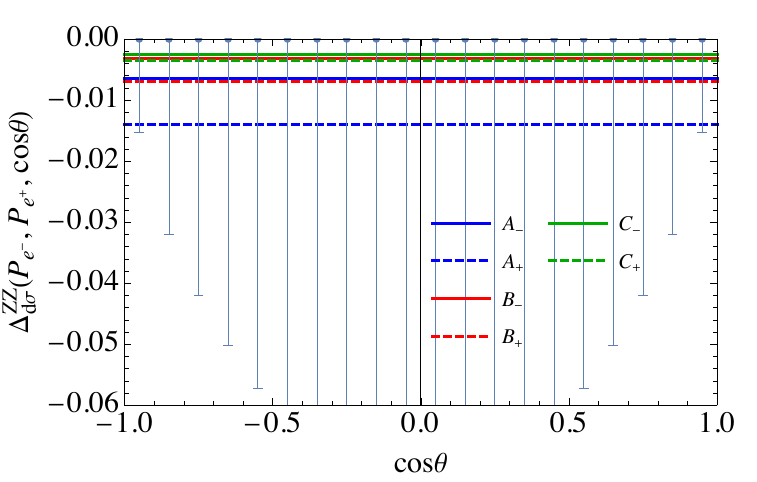}\\
\hline
\end{tabular}
 \caption{\small
 The deviations from the SM in the GHU for the $e^-e^+\to ZZ$
 process 
 $\Delta_{d\sigma}^{ZZ}(P_{e^-},P_{e^+},\cos\theta)$
 are shown with
 $(P_{e^-},P_{e^+})=(-0.8,+0.3),(0,0),(+0.8,-0.3)$
 for upper, middle and lower rows, respectively,
 where 
 $\Delta_{d\sigma}^{ZZ}(P_{e^-},P_{e^+},\cos\theta):=
 \left[\frac{d\sigma^{ZZ}}{d\cos\theta}
 (P_{e^-},P_{e^+})\right]_{\rm GHU}
 \left[\frac{d\sigma^{ZZ}}{d\cos\theta}
 (P_{e^-},P_{e^+})\right]_{\rm SM}^{-1}-1$.
 The left and right side figures show the deviation at
 $\sqrt{s}=250$\,GeV and $500$\,GeV, respectively. 
 The error bars represent statistical errors in the SM 
 at $\sqrt{s}=250 \,$GeV with 1\,ab$^{-1}$ data 
 and  at $\sqrt{s}=500\,$GeV with 2$\,$ab$^{-1}$ data 
 by using leptonic decays $Z\to \ell^+\ell^-$.
 Each bin is given by $\cos\theta=[k-0.05,k+0.05]$ 
 ($k=-0.95,-0.85,\cdots,0.95$).
 Note that $\mbox{Br}(Z\to \ell^+\ell^-)=(3.3658\pm 0.0023)$\% 
 \cite{ParticleDataGroup:2022pth}.
 }
 \label{Figure:Delta-dsigma-ZZ}
\end{center}
\end{figure}

In Figure~\ref{Figure:Delta-dsigma-ZZ}, the deviations from the SM  in
the GHU with polarized $e^\pm$ beams are shown.
The deviation from the SM in the GHU
model with unpolarized and left- and right-handed polarized $e^\pm$ beams
$(P_{e^-},P_{e^+})=(0,0),(-0.8,+0.3),(+0.8,-0.3)$.
Figure~\ref{Figure:Delta-dsigma-ZZ} shows that the deviation from the SM
is small for all the parameter sets.
The statistical errors in the SM are estimated by 
by using leptonic decays $Z\to \ell^+\ell^-$.
Since $\mbox{Br}(Z\to \ell^+\ell^-)=(3.3658\pm 0.0023)$\%
\cite{ParticleDataGroup:2022pth},
the branching ratio of 
$ZZ\to \ell^-\ell^+\ell^{\prime -}\ell^{\prime +}$ is 1.019\%.
It can be seen that while the decay to charged leptons can be measured
precisely, the statistical errors are overwhelmingly insufficient to see
the deviation from the SM in the GHU model. It may be possible to see
the deviation of the GHU model from the SM by using the decay modes of
the $Z$ bosons to hadrons, where the branching ratio of the $Z$ bosons
to hadrons is about 70\% and the branching ratio of the $Z$ bosons
to leptons and hadrons is about 10\%.
Therefore, it may be possible to explore up to the KK mass scale beyond
current experimental limits if the systematic errors in the decay of $Z$
bosons into hadrons can be sufficiently suppressed.

\begin{figure}[htb]
\begin{center}
\includegraphics[bb=0 0 363 232,height=5cm]{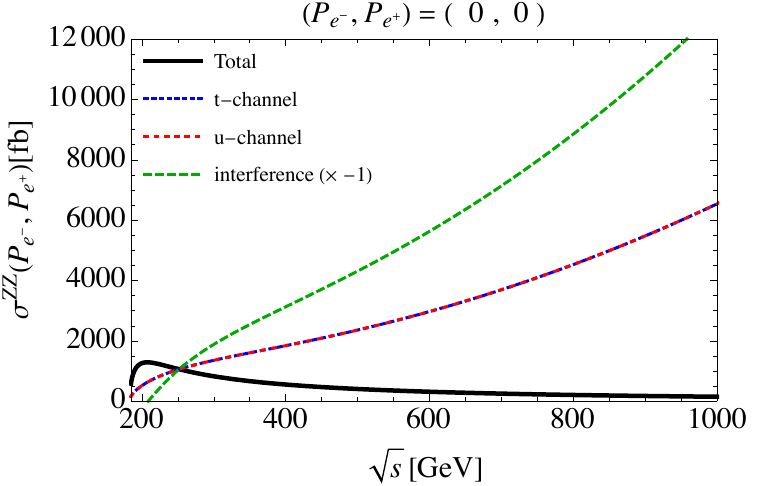}
\includegraphics[bb=0 0 363 236,height=5cm]{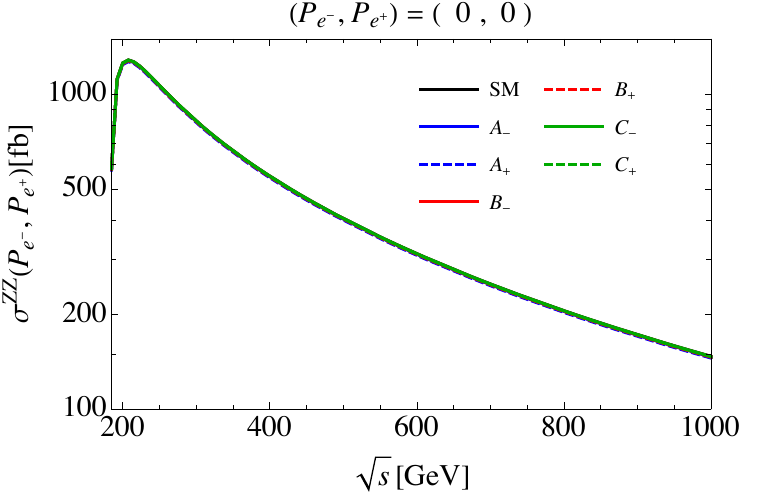}
 \caption{\small
 The total cross sections of the $e^-e^+\to ZZ$ process
 in the SM and the GHU model 
 $\sigma^{ZZ}(P_{e^-},P_{e^+})$ 
are shown in range of
 $\sqrt{s}=[185,1000]$\,GeV.
 The left figure shows
 the $\sqrt{s}$ dependence of 
 $\sigma^{ZZ}(P_{e^-}=0,P_{e^+}=0)$ in the SM 
 with unpolarized electron and positron beams, 
 Total stands for differential cross section including all the
 contribution from s-channel, t-channel, and interference terms.
 s-channel, t-channel, and interference stand for
 for cross section only including each contribution.
 The right figure shows the $\sqrt{s}$ dependence of 
 $\sigma^{ZZ}(P_{e^-},P_{e^+})$ in the SM and the GHU model whose
 parameters set are $A_\pm$, $B_\pm$, $C_\pm$.
 }
 \label{Figure:sigma-ZZ-low}
\end{center}
\end{figure}

From Figure~\ref{Figure:sigma-ZZ-low}, 
the total cross sections of the $e^-e^+\to ZZ$ process
in the SM and the GHU model are shown in range of
$\sqrt{s}=[185,1000]$\,GeV.
The left figure shows the $\sqrt{s}$ dependence of 
$\sigma^{WW}(P_{e^-}=0,P_{e^+}=0)$ in the SM 
with unpolarized electron and positron beams, 
Total stands for differential cross section including all the
contribution from t-channel, u-channel, and interference terms.
t-channel, u-channel, and interference stand for
for cross section only including each contribution.
The right figure shows the $\sqrt{s}$ dependence of 
$\sigma^{WW}(P_{e^-},P_{e^+})$ in the SM and the GHU model whose
parameters set are $A_\pm$, $B_\pm$, $C_\pm$.
From the left figure, we can see that for larger $\sqrt{s}$, stronger
cancellation between t-channel, u-channel, and interference terms is
taking place. This figure shows the SM case, but the same phenomenon
occurs in the GHU model.
From the right figure, we find that the cross sections in the GHU with 
the parameter sets $A_\pm$, $B_\pm$, $C_\pm$ are almost the same as that
in the SM.

\begin{figure}[htb]
\begin{center}
\includegraphics[bb=0 0 363 241,height=3.4cm]{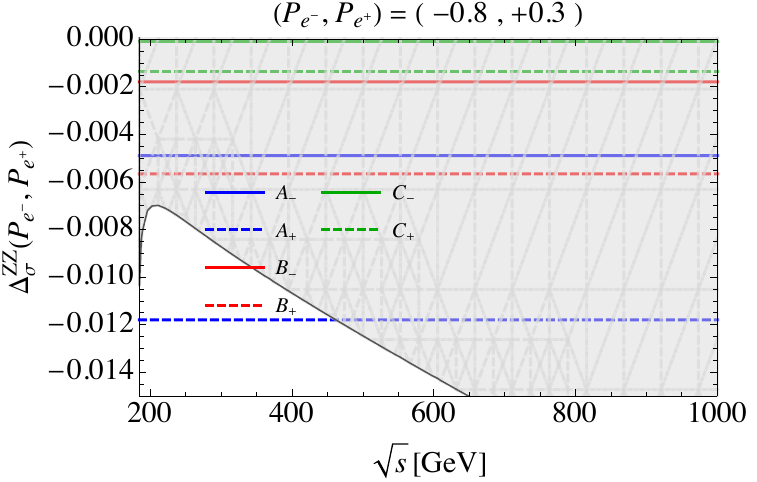}
\includegraphics[bb=0 0 363 241,height=3.4cm]{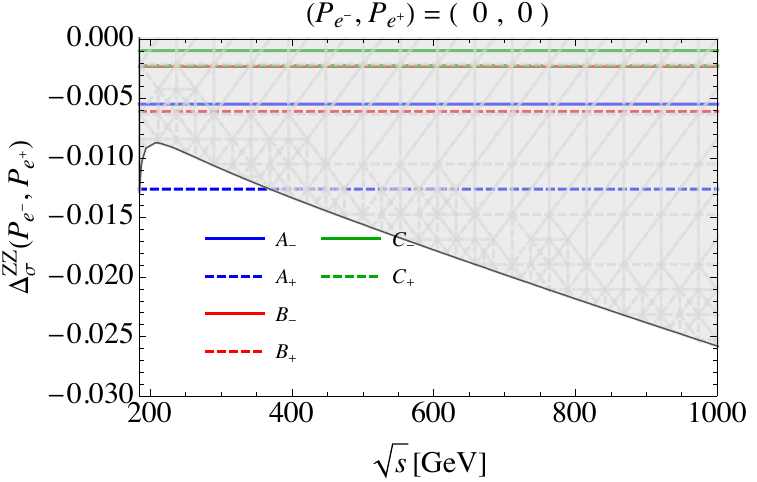}
\includegraphics[bb=0 0 363 241,height=3.4cm]{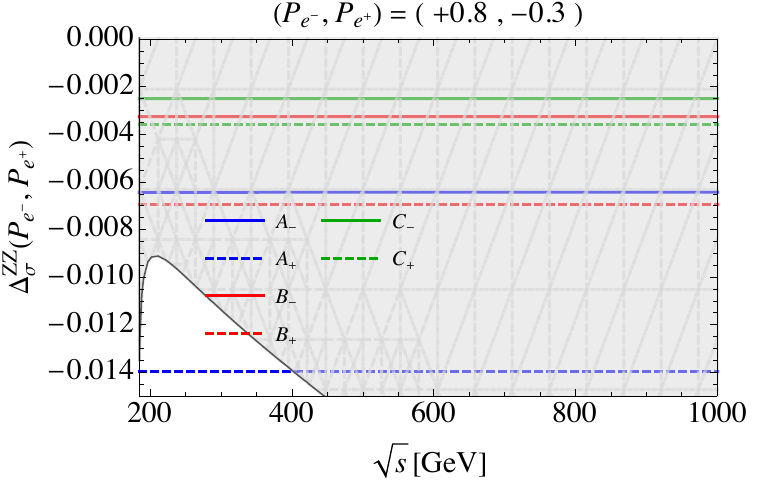}
 \caption{\small
 The $\sqrt{s}$ dependence of the deviation from the SM in
 the GHU models for the total cross sections,
 $\Delta_{\sigma}^{ZZ}(P_{e^-},P_{e^+})$,
 is shown for $(P_{e^-},P_{e^+})=(-0.8,+0.3),(0,0),(+0.8,-0.3)$,
 respectively,
 where 
 $\Delta_\sigma^{ZZ}(P_{e^-},P_{e^+}):=
 \left[\sigma^{ZZ}(P_{e^-},P_{e^+})\right]_{\rm GHU}/
 \left[\sigma^{ZZ}(P_{e^-},P_{e^+})\right]_{\rm SM}-1$.
 The gray region represents the $1\sigma$ statistical error
 estimated by using the total cross section of $Z$ boson pair
 production and  integrated luminosity  $L_{\rm int}=1\,\mbox{ab}^{-1}$.
 } 
 \label{Figure:Delta-sigma-ZZ}
\end{center}
\end{figure}

In Figure~\ref{Figure:Delta-sigma-ZZ}, the $\sqrt{s}$
dependence of the deviation from the SM in  the GHU models for the total
cross sections, $\Delta_{\sigma}^{ZZ}(P_{e^-},P_{e^+})$, is shown, where
$\Delta_\sigma^{WW}(P_{e^-},P_{e^+})$ is given in
Eq.~(\ref{Eq:Delta_sigma-ZZ}).
The statistical errors in the SM are estimated by 
by using leptonic decays $Z\to \ell^+\ell^-$.
Since $\mbox{Br}(Z\to \ell^+\ell^-)=(3.3658\pm 0.0023)$\%
\cite{ParticleDataGroup:2022pth},
the branching ratio of 
$ZZ\to \ell^-\ell^+\ell^{\prime -}\ell^{\prime +}$ is 1.019\%.
The decay mode of $Z$ bosons into charged leptons does not produce a
sufficient number of events.
If the decay of $Z$ bosons into hadrons can be used with sufficient
abundance and precision, it is possible that this process could be used
to explore well beyond the limits from the LHC experiment into the
$m_{\rm KK}$ region.

\begin{figure}[htb]
\begin{center}
\includegraphics[bb=0 0 363 240,height=5cm]{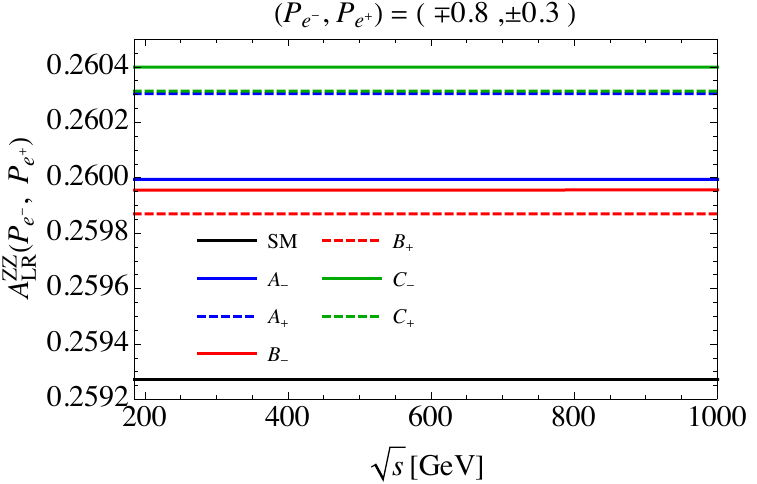}
\includegraphics[bb=0 0 363 235,height=5cm]{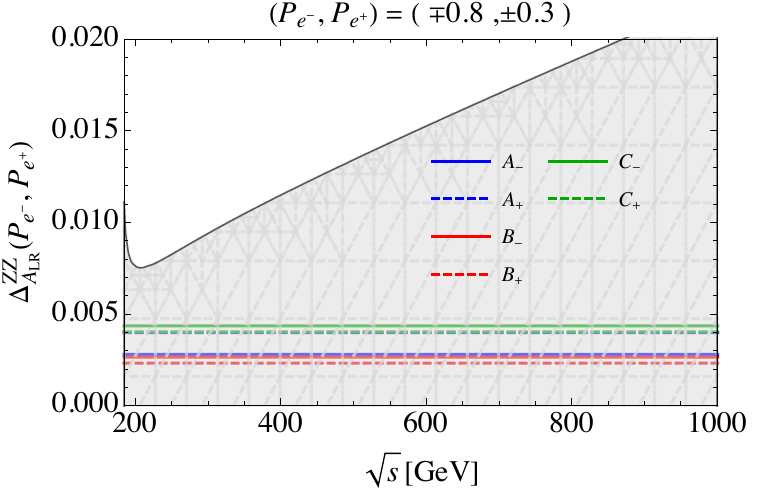}
 \caption{\small
 The $\sqrt{s}$ dependence of the left-right asymmetry of 
 the $e^-e^+\to ZZ$ process and the deviation from the SM  are shown. 
 The left figure shows the $\sqrt{s}$ dependence of
 $A_{LR}^{ZZ}(P_{e^-},P_{e^+})$ in the SM and the GHU model.
 The right figure shows the $\sqrt{s}$ dependence of
 $\Delta_{A_{LR}}^{ZZ}(P_{e^-},P_{e^+})$,
 where 
 $\Delta_{A_{LR}}^{ZZ}(P_{e^-},P_{e^+}):=
 \left[A_{LR}^{ZZ}(P_{e^-},P_{e^+})\right]_{\rm GHU}/
 \left[A_{LR}^{ZZ}(P_{e^-},P_{e^+})\right]_{\rm SM}-1$.
 The energy ranges $\sqrt{s}$ in the first and second figures are
 $\sqrt{s}=[200,3000] \,$GeV, $\sqrt{s}=[200,1000] \,$GeV, respectively.
 The gray region represents the $1\sigma$ statistical error in the SM 
 at each $\sqrt{s}$ with $1$$\,$ab$^{-1}$ for 
 each polarized initial states
 $(P_{e^-},P_{e^+})=(\mp0.8,\pm0.3)$
 by using the total cross section of $W$ boson pair production.
 } 
 \label{Figure:ALR-ZZ}
\end{center}
\end{figure}

In Figure~\ref{Figure:ALR-ZZ},
the $\sqrt{s}$ dependence of the left-right asymmetry $A_{LR}^{ZZ}$
of the $e^-e^+\to ZZ$ processes and the deviation from the SM
$\Delta_{A_{LR}}^{ZZ}$ are shown,
where $A_{LR}^{ZZ}$ and $\Delta_{A_{LR}}^{ZZ}$ are 
given in Eqs.~(\ref{Eq:ALR-ZZ-def}) and (\ref{Eq:Delta_A_LR-ZZ}),
respectively.
The $1\sigma$ statistical error in the SM
at each $\sqrt{s}$ with $1$$\,$ab$^{-1}$ for 
each polarized initial electron and positron
$(P_{e^-},P_{e^+})=(\mp0.8,\pm0.3)$
is estimated  by using leptonic decays $Z\to \ell^+\ell^-$.
As the same as the total cross section, the decay mode of $Z$ bosons
into charged leptons does not produce a sufficient number of events.
If the decay of $Z$ bosons into hadrons can be used with sufficient
abundance and precision, it is possible that this process could be used
to explore well beyond the limits from the LHC experiment into the
$m_{\rm KK}$ region.

\section{Summary and discussions}
\label{Sec:Summary}

In this paper we investigated the $W$ and $Z$ boson
pair production processes in the $SU(3)_C\times SO(5)_W\times U(1)_X$
GHU model.
First, by using the asymptotic behavior of the cross sections for large
$\sqrt{s}$, we derived the conditions given in
Eqs.~(\ref{Eq:Unitality-condition-1}) and
(\ref{Eq:Unitality-condition-2})
under which the $O(s)$ and $O(1)$ terms of the cross section for the 
$e^-e^+\to W^-W^+$ process cancel.
It is well-known that these conditions are satisfied in the SM,
but it is not obvious that the conditions are also satisfied in the GHU
model because of the deviation of the coupling constants in the GHU
model from those in the SM.
We confirmed that even in the GHU model 
not only the condition in Eq.~(\ref{Eq:Unitality-condition-1}) 
related to the unitality bound,
but also the condition in Eq.~(\ref{Eq:Unitality-condition-2}) 
are satisfied with very good accuracy.
Therefore, the unitality bound is satisfied.

Next, we found that 
from Figure~\ref{Figure:Delta-sigma-WW}
the deviation of the total cross section for the
$e^-e^+\to W^-W^+$ process from the SM in the GHU model with the
parameter sets $A_\pm$, which are consistent with the current
experimental constraints 
$(m_{\rm KK}\geq 13\,\mbox{TeV}, \theta_H\leq 0.10)$,
is about 0.5\% to 1.5\% and 0.6\% to 2.2\% for $\sqrt{s}=250$\,GeV and
500\,GeV, respectively, depending on the initial electron and  positron
polarization. 
To estimate whether these deviation is actually observable or not, we
estimated the statistical uncertainty by using the decay mode of $W^\pm$
decays into leptons. 
As a result, we found that it is possible to observe deviations from
the SM in the GHU model with parameter sets whose KK mass scale is
larger than 13\,TeV.

From an analysis similar to the $e^-e^+\to W^-W^+$ process,
we also found that for the $e^-e^+\to ZZ$ process the deviation from
the SM in the GHU model is at most 1\%.
To estimate whether these deviation is actually observable or not, we
estimated the statistical uncertainty by using the decay mode of $Z$
bosons into charged leptons.
Unfortunately, it is difficult to observe the deviation of cross
sections from the SM in the GHU model by using the $e^-e^+\to ZZ$
process because there is not a sufficient number of events.

In the GHU model, there is the large parity violation in the coupling
constants of $W'$ and $Z'$ bosons to quarks and leptons.
The bulk mass of the lepton determines whether the right-handed or
left-handed coupling 
constant is larger. Therefore, the deviation of cross sections from the
SM in the GHU model is expected to strongly depend on the initial
polarization of electrons and positrons.

Further theoretical and experimental studies in the GHU model are
necessary. As for theoretical studies, since the present study is a Born
level analysis, there are various issues to be addressed to give more
precise predictions, for example, off-shell final state contributions,
initial-state radiation, QCD corrections, etc.
Here we comment on the corrections to the $e^-e^+\to W^-W^+$ process.
The contributions of 1-loop corrections for the $e^-e^+\to W^-W^+$ process
in the SM and additional contributions from the $e^-e^+\to W^-W^+\gamma$ 
whose $\gamma$ is not detected are discussed in Ref.~\cite{Beenakker:1996kt}.
 These effects lead to an $O(10)\%$ contribution to the  $e^-e^+\to W^-W^+$ process at tree level in the SM.
 Since the coupling constants in the GHU model are almost identical to those in the SM, this subleading contributions are expected to be almost the same as those in the SM.
Therefore, the contributions to quantities defined from the ratio of
cross sections, etc. in the SM and the GHU model such as
$\Delta_{d\sigma}^{WW}$ and $\Delta_{\sigma}^{WW}$ 
in Figures~\ref{Figure:Delta-dsigma-WW} and \ref{Figure:Delta-sigma-WW}, 
in Figures~4 and 6, for example, is expected to be small.

\section*{Acknowledgments}

This work was supported in part
by the Ministry of Science and Technology of Taiwan under
Grant No. MOST-111-2811-M-002-047-MY2 (N.Y.).

\bibliographystyle{utphys}
\bibliography{../../arxiv/reference}

\end{document}